\documentclass[11pt]{article}
\pdfoutput=1
\usepackage{fancybox,amssymb,
graphicx,graphics,color,epsfig,subfigure,latexsym,tabularx,times,rotating,amsmath,ctable,multirow,
verbatim,setspace, array,pict2e}
\usepackage{subfigure}
\PassOptionsToPackage{hyphens}{url}\usepackage{hyperref}
\usepackage{natbib}
\usepackage{lscape}
\usepackage{fancyvrb}
\usepackage[T1]{fontenc}
\usepackage[sc]{mathpazo}
\DeclareMathOperator*{\argmin}{arg\,min}

\usepackage{amsmath}
\usepackage{amsfonts}
\usepackage{amssymb}
\newcommand{\bzero}{\mathbf{0}}

\newcommand{\be}{\mathbf{e}}
\newcommand{\bx}{\mathbf{x}}
\newcommand{\bX}{\mathbf{X}}
\newcommand{\by}{\mathbf{y}}
\newcommand{\bY}{\mathbf{Y}}
\newcommand{\bbeta}{\boldsymbol{\beta}}
\newcommand{\beps}{\boldsymbol{\epsilon}}
\newcommand{\rss}{\text{\it RSS}}
\newcommand{\mv}{[-v]}
\newcommand{\mt}{[-t]}
\newcommand{\mg}{[-g]}
\newcommand{\mivph}{\overline{\ivph}}
\newcommand{\mbvph}{\overline{\bvph}}
\newcommand{\ivph}{Z}
\newcommand{\svph}{Z}
\newcommand{\stvph}{\widetilde{Z}}
\newcommand{\bvph}{\mathbf{Z}}
\newcommand{\btvph}{\mathbf{\tilde{Z}}}
\newcommand{\beff}{\omega}
\newcommand{\sumvar}{s}
\newcommand{\ivar}{W}
\newcommand{\svar}{W}
\newcommand{\bvar}{\mathbf{W}}
\newcommand{\btvar}{\mathbf{\tilde{W}}}
\newcommand{\meff}{\nu}
\newcommand{\bmeff}{\boldsymbol\nu}
\newcommand{\vareff}{\omega_{\svar}}
\newcommand{\sumgrp}{s}
\newcommand{\igrp}{G}
\newcommand{\sgrp}{G}
\newcommand{\bgrp}{\mathbf{G}}
\newcommand{\btgrp}{\mathbf{\tilde{G}}}
\newcommand{\geff}{\nu}

\newcommand{\grpeff}{\omega_{\sgrp}}
\providecommand{\abs}[1]{|#1|}

\newcommand{\pr}{\text{Pr}}
\newcommand{\expect}{\mathbb{E}}

\newcommand{\gauss}{\mathcal{N}}

\newcommand{\simiid}{\stackrel{\text{iid}}{\sim}}
\newcommand{\tunif}{\text{Unif}}
\newcommand{\tbern}{\text{Bernoulli}}
\newcommand{\tgamma}{\text{Gamma}}
\newcommand{\tbeta}{\text{Beta}}

\newcommand{\hbfdr}{\widehat{\text{BFDR}}}
\newcommand{\grpsub}{\genfrac{}{}{0pt}{}{v:\gamma(v)=g,}{\ivph_v=1}}
\newcommand{\cmax}{C_{\text{max}}}

\bibpunct{(}{)}{;}{a}{}{,}

\title{Genetic variant selection: learning across traits and sites}
\author{Laurel Stell\textsuperscript{1} and Chiara Sabatti\textsuperscript{2}}

\begin{document}

\maketitle

{%\centering% adjust the vertical skips according to your needs
\vspace*{-0.3cm}
\noindent\textsuperscript{1} Department of Health Research and Policy,  Stanford University,
  Stanford, CA 94305, U.S.A.\\
\textsuperscript{2} Departments of Health Research and Policy and Statistics, Stanford
  University, Stanford, CA~94305, U.S.A.

}

\vspace{1em}

\begin{abstract}
We consider resequencing studies of associated loci and the problem of
prioritizing sequence variants for functional follow-up. Working within the
multivariate linear regression framework helps us to account for the joint effects of multiple genes; and adopting a Bayesian approach  leads to posterior
probabilities that coherently incorporate all information about the variants' function. We
describe two novel prior distributions that facilitate learning the role of each
variable site by borrowing evidence across phenotypes and across mutations in the same
gene. We illustrate their potential advantages with simulations and re-analyzing
a dataset of sequencing variants.
\end{abstract}

\section{Introduction}

Genome-Wide Association Studies (GWAS) have
% For publication, use next line instead of the preceding line.
%\lettrine[lines=2]{\color{color2}G}{}enome-Wide Association Studies (GWAS) have
allowed human geneticists to compile a rather long list of loci where DNA
variation appears to be reproducibly associated to phenotypic variability
\citep{GWAS}. While these might represent only a subset of the portion of the
genome that is important for the traits under study \citep{MetV09}, there is
little doubt that understanding the characteristics and mechanisms of functional
variants at these loci is a necessary next  step.  As resequencing becomes ever
more affordable, follow-up investigations of GWAS loci often start with a
comprehensive catalogue of their genetic variants in a sample of thousands of
individuals, raising the question of how to sort through these results.

Among the many challenges,
let us discuss two. First, common variants are often correlated and it is
difficult to distinguish their roles without accounting for the broader genetic
background of the individuals who carry them. Second, rare variants
are present in a small enough portion of the sample that statistical statements
become impossible.  With this in mind, it has been noted that  (a)~it is important to 
account for correlation between variants to obtain useful ranking; (b)~we should
increasingly be able to take advantage of the information gathered through other
studies; and (c)~Bayesian models provide a principled approach to guide variant
prioritization. To adequately select among variants in the same locus (defined
as a genomic region that might encompass multiple genes but that corresponds to
the same association signal in a GWAS study),
researchers have resorted to model selection approaches \citep{VetH12} or
approximations of the joint distribution of univariate test statistics
\citep{FetBS13, HetE14}. Prior information on variant annotation has been
incorporated in models for eQTL \citep{VetP08} and more recently for general
traits \citep{P14,KetP14,CetZ14}, and annotation programs increasingly attempt
to include information on identified genetic loci \citep{ANNOVAR}.
Prioritization often relies on Bayes's theorem, and Bayesian methods have
received renewed attention in the context of GWAS data analysis
\citep{GS2011,peltola1step,PeltolaAdapt}, genomic prediction \citep{G2013}, and  the evaluation of heritability
\citep{ZetS13}. 

In this context, we  explore the advantages of
a careful specification of the prior distributions on variants, by allowing
sharing of information across multiple phenotypes and across neighboring rare
variants. We are motivated
by the analysis of an exome resequencing study \citep{SetF14} in which individual level data is available for exomic variants  at multiple genomic loci that have demonstrated evidence in GWAS of association to lipid traits.
By design, the vast majority of measured variants is coding or in UTR,
that is, in portions of the genome with high prior probability of harboring
functional mutations.  Annotation can help distinguish the role of synonymous
variants, and conservation scores can be used to predict the effect of nonsynonymous
ones; but annotation cannot be used to discount the importance of a large number
of non-coding variants that one can expect to occur in a whole genome sequencing dataset.
Measures on levels of High Density Lipoprotein  (HDL), Low Density
Lipoprotein  (LDL), and triglycerides (TG) are available for the study subjects,
and we are interested in capitalizing on the multidimensional nature of the
phenotype. Prior analyses of this dataset  \citep{SetF14, BetC15} have
illustrated the importance and the challenges of multivariate linear models, and we explore here the advantages offered by carefully selecting priors for Bayesian models. Abstracting from the specifics of this dataset, we show how hierarchical prior distributions  can be adapted to learn about the
functionality of a variant  by (i) looking across multiple phenotypes and (ii)
aggregating the effects of multiple rare variants in the same gene. Since the
power of Bayesian methods in borrowing information is well known, it is not
surprising that others have explored their application in this context. For
example, \cite{YetL11} illustrate the use of priors to model a group effect for
multiple rare variants, while \cite{S2013} describes models for the analysis of
multiple traits. Our approach, however, is distinct from others in that  it
strives to achieve all of the following: (1)~constructing a multivariate linear
model that simultaneously accounts for the contributions of multiple genes and
genomic loci; (2)~providing inference on variant-specific effects---while
linking information across traits and genomic sites;
and (3)~accounting for the large number of variants tested,
effectively enacting a form of multiple comparison adjustment. 

 This paper is organized as follows. We devote
Section~\ref{sec:priors} to introducing the novel priors in the context of the 
 genetic model, using an approximation of the posterior distribution to illustrate
their inferential implications. Section~\ref{sec:mcmc} describes the MCMC
scheme used to sample the posterior,   the setting used for simulations, and the
criteria for comparison of methods.  Section~\ref{sec:results} presents the
results of simulation studies highlighting the potential of our proposal, as
well as the description of the analysis of the motivating dataset. 

\section{Prior distributions on genetic variants}
\label{sec:priors}

One characteristic of a genetic study based on resequencing, as contrasted to genotyping, is that researchers aim to collect a comprehensive catalogue of all genetic variants. This has implications for the statistical models used to analyze the data and the prior assumptions.
Let $n$ be the number of subjects in the study and $p$ the number of polymorphic sites assayed. We will use $y_i$ to indicate the phenotypic value of subject $i$ and $X_{iv}$ the genotype of this subject at variant $v$ (typically coded as minor allele count). The simplest genetic model for a heritable phenotype is of the form
% LLS changed eta to zeta because eta is correlation between predictor and trait
$$y_i=\sum_{k\in\mathcal{T}}G_{ik}+\zeta_i,$$
where $\zeta_i$ incapsulate all nongenetic effects and $G_{ik}$ for
$k\in\mathcal{T}$
represent the contributions of a set $\mathcal{T}$ of genes that act additively
and independently.
Without loss of generality and following a standard practice in GWAS, we will
assume that the effects of nongenetic determinants of the phenotypes have been
regressed out from $y_i$ so that
$\zeta_i$ can be considered independent `error' terms.
Let us assume that the genetic effects are a linear function of minor allele counts so that
\begin{equation}
y_i=\sum_{v\in\mathcal{V}}\beta_vX_{iv}+\epsilon_i
\label{truth}\end{equation}
for a set $\mathcal{V}$ of causal variants with $\epsilon_i$ {\em iid}
$\gauss(0,1/\rho).$ 
Although this assumption is substantial, it only has the role of simplifying notation.
While \eqref{truth} represents the true genetic architecture of the trait,
the membership of $\mathcal{V}$ is unknown in a typical association study,
so the relation between the phenotype and genetic variants is expressed as
\begin{equation}
  y_i=\sum_{v=1}^p\beta_vX_{iv}+\epsilon_i, \hspace{2em}
       \epsilon_i \simiid \gauss\left(0,\frac{1}{\rho}\right),
                     \label{big}
\end{equation}
summing over all variable sites and with the understanding that only an
(unknown) subset of $\bbeta=(\beta_1, \ldots,\beta_p)$ is different from 0.
Below we will use the compact matrix notation $\by=\bX\bbeta+\beps$.
Using \eqref{big} to describe the relation between traits and genotypes depends
heavily on the assumption that a resequencing study assays all variants.
In GWAS, on the other hand, causal variants might be untyped, which means their
contributions are partially captured by correlated variants and partially
included in the error term.  It would still be meaningful in that context to use
a linear model to link phenotype and genotypes. However, in GWAS, the errors cannot be assumed
independent, and the interpretation of the coefficients of $\bX$---as well as
their prior distribution---is substantially more complicated. We note that mixed effects models can be used to address the first concern \citep{KetE10}.

The parameters in model \eqref{big} are $\bbeta$ and $\rho$; we now focus on their prior distribution.
Following standard practice, we  take
 $\rho\sim\tgamma(\alpha_{\rho},\lambda_{\rho})$. (See \cite{GS2011} for another
approach that specifically targets GWAS and relies on heritability information.) On the vector $\bbeta$, we
want a prior that reflects our model selection goals and our understanding of
the genetic architecture. There are several aspects to consider: (a)~given the exhaustive nature of the genotyping process, we believe that
most of the variants available do not directly influence the trait; (b)~it seems
reasonable that a variant that influences one trait (so that its effect size
is definitely not zero) might also influence other traits; and
finally (c)~it appears likely that if a rare variant influences the outcome,
other nearby rare variants might also have an effect. Our main goal is
to describe prior distributions on $\bbeta$ that incorporate these  beliefs.  We
start by recalling one class of priors that reflect (a) and then move on
to generalizations that account for the sharing of information implied by (b)
and (c). In what follows, we assume that the allele counts in the column  of $\bX$ have been standardized to have mean zero and variance one.

\subsection{Priors incorporating sparsity}

The prior belief that only a fraction of the typed variants has an effect on the phenotype is but one instance of what is a common assumption in high-dimensional statistics, {\em i.e.} that the parameter $\bbeta$ of interest is sparse. 
To specify a prior on $\bbeta$ that gives positive probability to vectors with a
number of coordinates equal to zero, we rely on a construction by \cite{gmgibbs} and introduce 
a vector of indicator variables $\bvph$ such that
$\ivph_v=0$ implies $ \beta_v=0$.   The $\ivph_v$ are  {\em iid}
Bernoulli with parameter $\beff$, which governs the sparsity of the model and
 has a Beta$(A_{\beff},B_{\beff})$ prior. Let
$\bbeta_{\svph}$ indicate the collection of elements of $\bbeta$ corresponding
to nonzero elements of $\bvph$, and let $\bX_{\svph}$ be the corresponding columns of
$\bX$. It has been found useful to assume
$(\bbeta_{\svph}|\bvph,\rho,\tau) \sim
\gauss\left(0,\frac{\tau^2}{\rho}\Sigma_{\svph}\right),$ where
$\Sigma_{\svph}$ is  a known matrix and $\tau \sim$ Unif$(\tau_1,\tau_2)$ links
the error variance to the size of the $\bbeta$ coefficients.  In the literature,
$\Sigma_{\svph}$ mainly has one of two forms: $I_{\abs{\bvph}}$ (the identity matrix of size $\abs{\bvph}$, where
$\abs{\bvph}$ indicates the number of nonzero components of the vector $\bvph$)
or $n(\bX_{\svph}^T\bX_{\svph})^{-1}$, which is referred to as the g-prior
\citep{Z86} (and is a viable choice only when $\abs{\bvph}<n$).
Various views on the choice of $\Sigma_{\svph}$ have been put forth
\citep{cgm,heaton,GS2011}, but the strongest argument for the g-prior is that it
provides computational benefits (see below).
For either choice of $\Sigma_{\svph}$, all of its diagonal entries are equal,
resulting in an equal prior variance for each of the $\beta_v$.  
Given the standardization of the columns of $\bX$, this implies that the original effect sizes are expected to be larger for rare
variants than for common variants, which is reasonable. 

% LLS says: \rho_t in this figure should be \rho
\begin{figure}[htbp]
\centering
\hspace*{-.25cm}
\includegraphics[scale=0.9]{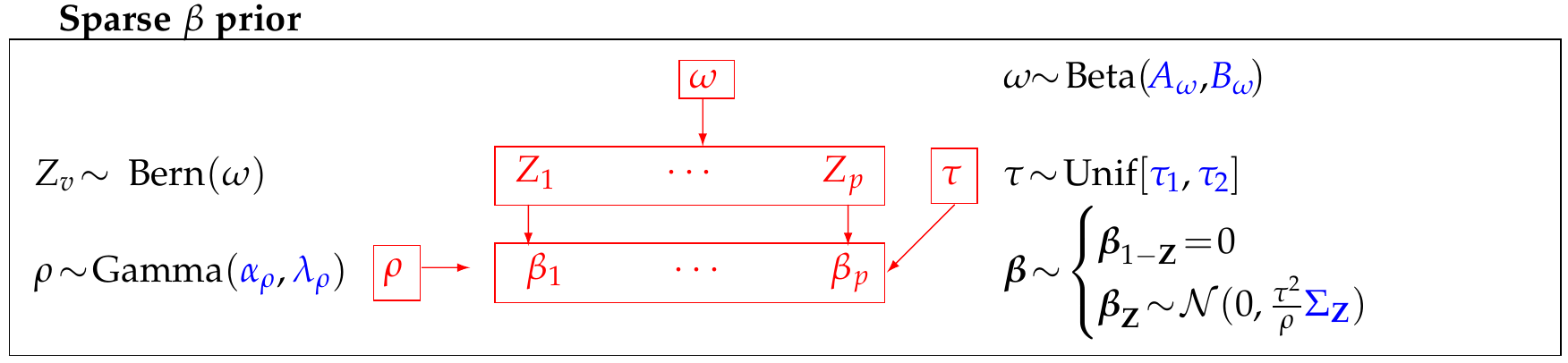}
\vspace*{-.3cm}
\caption{Schematic representation of the {\em Sparse}  prior distribution
on $\bbeta$. Hyperparameters are indicated in blue. The red portion describes
all random objects and their dependency structure; unless explicitly indicated,
 random variables are independent. Boxes identify variables that share
one of the distributions  depicted in black.}\label{Sparse}
\end{figure}

One of the advantages of the prior summarized in Figure~\eqref{Sparse} is that
the derived posterior distribution can be analytically integrated with respect
to $\beff$, $\bbeta$, and $\rho$. While a MCMC is still needed to fully explore
the posterior and carry out inference, we can rely on a
collapsed Gibbs sampler that focuses only on $\tau$ and the indicator variables
$\bvph$. This reduces the computation at each
iteration and  improves its convergence rate \citep{liu1994}.  The prior
densities for $\tau$ and $\bvph$  will be denoted, respectively, as $f_\tau(\tau)$
and $f_{\svph}(\bvph)$---the latter being easily obtained from the beta-binomial
distribution assumed for $\abs{\bvph}$. As shown in the appendix, integrating
$\bbeta$ and $\rho$ out gives the marginal posterior density
\begin{equation}
  f_{\bvph,\tau}(\bvph,\tau|\by)
    \propto f_\tau(\tau) f_{\svph}(\bvph)
        \Big( \lambda_\rho + \frac{S_{\svph}^2}{2}
                     \Big)^{-\left(\frac{n}{2}+\alpha_\rho\right)}
            \frac{\det(\Omega_{\svph})^{1/2}}
                 {\tau^{\abs{\bvph}} \det(\Sigma_{\svph})^{1/2}},
                            \label{eq:basicpost}
\end{equation}
where
$\Omega_{\svph}^{-1} = \bX_{\svph}^T\bX_{\svph} + \tau^{-2} \Sigma_{\svph}^{-1}$
and
$S_{\svph}^2 = \by^T \by - \by^T \bX_{\svph} \Omega_{\svph} \bX_{\svph}^T \by$.
Choosing $\Sigma_{\svph}$ as in the g-prior leads to a simplification of the
ratio in \eqref{eq:basicpost},
thereby avoiding the evaluation of one determinant at each iteration.

Despite the need to evaluate numerically interesting summaries of the posterior
of $\bvph$, we obtained an approximation (whose derivation and applicability is described in the
appendix) to gain a general understanding of how hyperparameters
and data contribute to the final inferential results.
Specifically, we focus on the
posterior expected value of $\svph_v$, indicator of variant $v$, conditional on
the indicators of all other variants $\svph_{[-v]}$. In the case of orthogonal regressors, this expectation can be approximated as
%% It took me a bit of work to avoid overfull hbox here.
%In the following, $\bvph_{\mv}$ denotes the subvector of $\bvph$ obtained by
%removing $\ivph_v$. 
% Along the lines of \cite{MalW11}, we present  an approximation of the posterior
%expected value $\expect[\ivph_v|\bvph_{\mv},\tau,\by]$,
%which---while derived under somewhat
%restrictive assumptions---illustrates the role of various elements of the prior
%in the final inferential result.  Consider the case when the columns of $\bX$
%are orthogonal, which implies $\bX^T\bX\approx nI_p$ and the two choices of
%$\Sigma_{\svph}$ are essentially the same.  Furthermore,
%$\langle \bx_v, \by \rangle = \langle \bx_v, \bX \bbeta + \beps \rangle
%    \approx n \beta_v + \langle \bx_v, \beps \rangle$,
%which is distributed as
%$\gauss(0,\frac{n}{\rho} (\ivph_v n \tau^2 + 1) )$; in this context, distinguishing signal from
%noise requires $n\tau^2 \gg 1$, so we assume this to be the case.
%We also assume that the portion of variance explained by the model is small, as
%is typically the case for genetic association.  If one further chooses
%$\alpha_\rho=\lambda_\rho \ll \frac{n}{2}$, then 
\begin{equation}
  \expect[\ivph_v|\bvph_{\mv},\tau,\by]^{-1}
    \approx  1 + \tau \sqrt{n}
                        \frac{B_{\beff}+p-|\bvph_{\mv}|-1}
                             {A_{\beff}+|\bvph_{\mv}|}
                        \big(1-\eta_v^2\big)^{n/2},
                               \label{eq:basicEgammaj}
\end{equation}
where $\eta_v = \bx_v^T\by/\sqrt{n \by^T\by}$ is approximately the correlation
between variant~$v$ and the trait.
From \eqref{eq:basicEgammaj}, one gathers that  increasing
$\abs{\bvph_{\mv}}$, which is the number of variants already used to explain the trait,
increases the chance of an additional variant $v$ to be considered relevant. This is a consequence of the fact that the parameter $\omega$, which describes the sparsity of $\bbeta$ and hence the degree of poligenicity of the trait,  is learned from the data (rather than set at a predetermined value). When a large number of variants have been found relevant, the trait is estimated to be highly poligenic and hence it is judged more likely that an additional variant might contribute to its variability.
  On the
other hand, augmenting the total number of genotyped sites $p$ will make it
harder for any specific variant $v$ to be judged important; this is  adjusting
for the look-everywhere effect, an important step  in gene mapping studies.

Now that we have introduced this basic framework, we can consider
modifications that facilitate learning about the role of a variant across
multiple traits and in the context of neighboring sites. We start with the first problem. 

\subsection{Learning across traits}
\label{multi}

One of the characteristics of current genetic datasets is the increased availability of
multidimensional phenotypes. This is  due partly to the automation with which
many traits are measured and partly to the increased awareness that precise phenotypic measurements are needed to make progress in our understanding of the underlying biological pathways. Having records of multiple traits in the same dataset allows for cross-pollination of genetic information.
On the one hand, if a genetic variant is functional, it can
be expected to impact more than one phenotype. On the other hand, even if noise
in one phenotype makes it hard to distinguish the predictive power of a causal
variant from that of a non-causal neighboring variant, it is much less likely
that multiple traits would have noise correlated in such a way that causal and
non-causal variants are indistinguishable for all of them.  With this in mind,
let us generalize the variant selection problem described in the previous
section to handle multiple traits. 

Extending the notation, let $\by_t$ be the standardized values for trait~$t$,
$\bbeta_t$ the coefficients of $\bX$ in the mean $\expect[\by_t]$, and $\bvph_t$
the corresponding indicator vector.   We organize these by column in a
$n\times q$ matrix $\bY$, a $p\times q$ matrix $\bbeta$, and a $p\times q$ matrix
$\bvph$.  Also let $\sumvar_v = \sum_t \ivph_{vt}$ denote the number of traits
associated with variant~$v$, let $\bbeta_{\bvph_t}$ be the entries of
$\bbeta_t$ corresponding to entries equal to one in $\bvph_t$, and let
$\bX_{\bvph_t}$ be the corresponding columns of $\bX$.  The
data-generating model is $\by_t = \bX_{\bvph_t} \bbeta_{\bvph_t} + \beps_t$ with
$\beps_t \sim \gauss\left(0,\frac{1}{\rho_t}I_n\right)$, and the priors on
$\rho_t$ are simple extensions of the one used previously: $ \rho_t$ {\em iid}
$\tgamma(\alpha_\rho,\lambda_\rho)$.
Note that this model assumes that, conditionally on the genetic variants that
influence them, the traits are independent; specifically, there are no shared environmental
effects. This assumption might or might not be appropriate  depending on context, but
the  prior distribution on $\bbeta$ that we are about to describe can be used
also for models that do not rely on this assumption.

  We want a prior for $\bbeta$ that continues to enforce sparsity but that allows learning about the role of a variant
across traits. One possibility, first proposed by \cite{jia07}, is to  introduce a variant-specific
probability of functional effect $\meff_v$, constant across traits and a
priori independent with $\meff_v \sim \tbeta(A_v,B_v),$ where $A_v$ and $B_v$
can capture annotation information. Following the setup of the
previous section, we then take  $(\ivph_{vt}|\meff_v)$ independent
$\tbern(\meff_v) $, set $\beta_{vt}=0 $ whenever $\ivph_{vt}=0$, and let
$(\bbeta_{\bvph_t}|\bvph_t,\rho_t,\tau)$ be independent across $t$ with
distribution $\gauss(0,\frac{\tau^2}{\rho_t} \Sigma_{\bvph_t})$. As before,
$\tau \sim \tunif(\tau_1,\tau_2)$. 

As detailed in the appendix, we can derive an approximation analogous to
\eqref{eq:basicEgammaj}:
\begin{equation}
  \expect[\ivph_{vt}|\bvph_{[-(vt)]},\tau,\bY]^{-1}
    \approx  1 + \tau \sqrt{n}
                        \frac{B_v+q-\sumvar_{v,\mt}-1}{A_v+\sumvar_{v,\mt}}
                        \left(1-\eta_{vt}^2\right)^{n/2},
                               \label{eq:multiEgammaj}
\end{equation}
where $\eta_{vt} = \frac{\bx_v^T\by_t}{\sqrt{n \by_t^T\by_t}}$ and 
$\sumvar_{v,\mt} = \sum_{\ell\neq t}\ivph_{v\ell}$ tallies the number of
phenotypes for which the variant $v$ has been judged relevant. This highlights a
consequence of the selected prior distribution: as the total number of phenotypes $q$ here has taken the role of
$p$ in \eqref{eq:basicEgammaj},
 the role of each variant  is judged not in reference to
all the other variants but only in comparison to the effect of the same variant
across traits. In other words, while there is learning across phenotypes, there is no adjustment for the multiplicity
of queried variants.
\cite{bott11} previously observed that sparsity of $\bvph_t$ could not be
controlled by specification of the priors in this approach, and proposed letting
$\ivph_{vt}$ have Bernoulli parameter $\nu_v\omega_t$ with independent priors on
each factor.

We propose a different remedy by introducing another layer in the hierarchical
priors. Let $\bvar$ be a vector of indicator
variables of length $p$: if $\ivar_v = 0$, then  $\meff_v = 0$; if $\ivar_v=1$,
$\meff_v \sim \tbeta(A_v,B_v)$. We take $\ivar_v$ {\em iid} $\tbern(\vareff)$ with
$\vareff \sim \tbeta(A_{\svar},B_{\svar})$;  the $(\ivph_{vt}|\meff_v)$ are independent
$\tbern(\meff_v)$, as before. 
% LLS was here. There were two different attempts (commented out below) to
% explain this prior, so I've smooshed them together.
\begin{figure}[htbp]
\centering
\hspace*{-.25cm}
\includegraphics[scale=0.9]{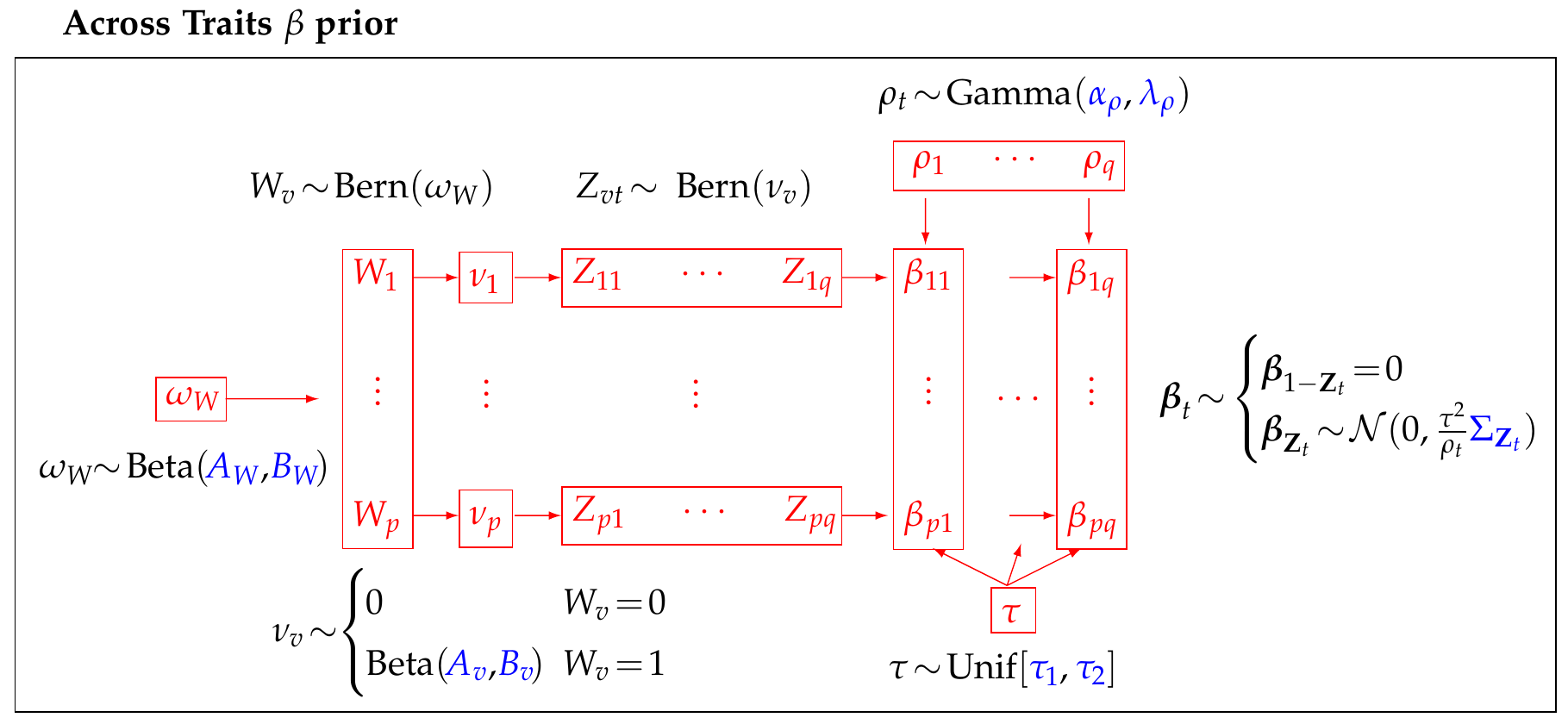}
\vspace*{-.3cm}
\caption{Schematic representation of the {\em Across Traits}  prior distribution
on $\bbeta$. Hyperparameters are indicated in blue. The red portion describes
all random objects and their dependency structure; unless explicitly indicated, random variables are independent. Boxes identify variables that share
one of the distributions  depicted in black.}\label{AT}
\end{figure}
The schematic in Figure~\ref{AT} summarizes this prior proposal.
The existence of a specific parameter $\nu_v$ for each site $v$ allows
variation in the average number of impacted traits per variant; some variants can
be highly
pleiotropic, while others are relevant for one trait only.   The sparsity
parameter $\beff_{\ivar}$ is once again estimated from the data, allowing for
multiplicity adjustment. The
introduction of $\bvar$ effectively specifies a hierarchical prior on
$\nu_1,\ldots, \nu_p$; 
among the many possible ways to accomplish this, the one we adopt emphasizes the role of the sparsity
parameter $\beff_{\ivar}$ and is easily interpretable.
The appendix presents an indicative approximation of the
posterior conditional expected values of $\ivar_v$ similar to
\eqref{eq:basicEgammaj} and \eqref{eq:multiEgammaj}. It depends on all phenotypes
(enabling learning across traits), but the total number of variants $p$ has
again become the leading factor for effective multiplicity correction.
We will refer to this prior as learning {\em Across Traits}.  We include
the first proposal in some comparison studies, indicating it as the {\em
Unadjusted} approach to emphasize the fact that it does not include an
adjustment for multiplicity.

This may be an appropriate point at which to clarify the relation between the prior we
are proposing and the traditional investigation of pleiotropy versus coincident
linkage. The latter terminology derives from linkage studies, where the nature
of the signal is such that the localization of variants with possible
pleiotropic effects is possible only up to a certain genomic interval. This interval
might contain one variant affecting multiple traits or contain different
variants, each affecting a subgroup of the traits. First, it is worth noting
that in this paper we are working in the context of association studies, which allow for a
much finer resolution than linkage studies.  The occurrence of multiple variants
affecting multiple traits within the same linkage disequilibrium block is less
likely, given that LD blocks are shorter than linkage regions. Secondly, ours is
a fixed effects model using sequence data.  We are aiming to estimate the
specific effect of each variant rather than simply identifying a locus with a
random effects model. Our framework, then, automatically considers two options:
one variant affecting multiple traits or multiple variants affecting separate
traits. The choice between these two alternatives is made on the basis of the
posterior probability of the two models. This being said, it is important to
recall that if two neighboring variants in LD affect two separate traits, the
posterior probabilities of the two alternative models might be similar. The
prior we introduce favors pleiotropy in the sense that it recognizes as likely
that some variants affect multiple genes, but it does not exclude the
alternative explanation, allowing the data to tilt the posterior in either
direction. We have investigated this with simulations in the supplementary
material.

\subsection{Learning across sites}

We now consider another form of `learning from the experience of others' to
improve our ability to identify functional  variants. We focus on rare variants, which are observed in just a handful of
subjects and for which it might be impossible to estimate individual effects. It
is reasonable to assume that if one rare variant in a gene has an impact on a
trait, other rare variants in the same gene also might be functional; with an appropriate hierarchical prior we might increase our ability to learn the effects of these variants.
Of course a similar assumption might also be reasonable for common variants, but
given that we observe these in a sufficiently large sample, we aim to estimate
their individual effect without convolving their signal with that of others.

\begin{figure}[htbp]
\centering
\hspace*{-.25cm}
\includegraphics[scale=0.9]{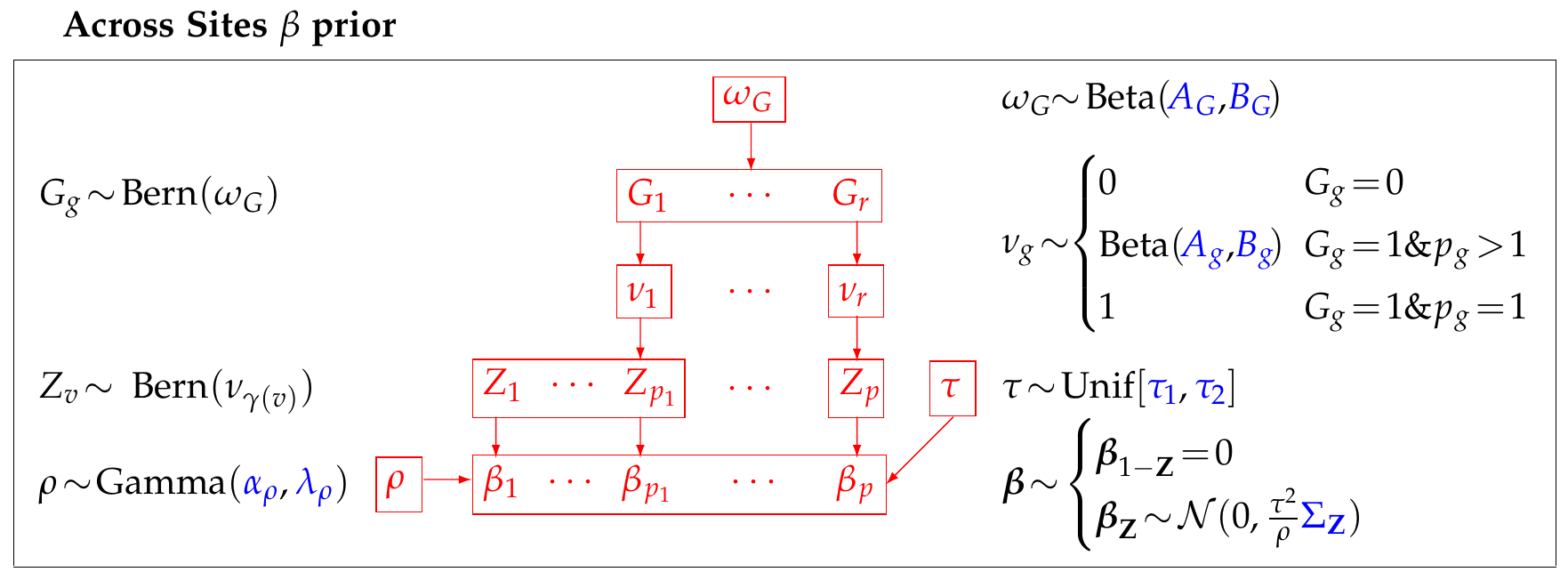}
\vspace*{-.3cm}
\caption{Schematic representation of the {\em Across Sites}  prior distribution
on $\bbeta$. Hyperparameters are indicated in blue. The red portion describes
all random objects and their dependency structure; unless explicitly indicated, random variables are independent. Boxes identify variables that share
one of the distributions  depicted in black.}\label{AS}
\end{figure}

The data-generating model is again \eqref{big}. 
We define $r$ groups of variants, and we use $\gamma(v)$ to indicate the group to
which variant~$v$ belongs.  Let $\bgrp=(\sgrp_1,\ldots,\sgrp_r)$ be a vector of
indicator variables associated to  the groups; we use these to link
information from different variants. Specifically, if $\sgrp_g = 0$, then the
proportion $\nu_g$ of causal variants in group $g$ is equal to zero;
otherwise, $\nu_g \sim \tbeta(A_g,B_g)$ (setting $\nu_g=1$ for groups
comprised of only one variant). The variant-specific indicators $\svph_v$ are
{\em iid} Bernoulli with parameter $\nu_{\gamma(v)}$.  Similarly to prior specifications, 
 $(\igrp_g|\grpeff)$ are {\em iid} $\tbern(\grpeff)$ with
$\grpeff \sim \tbeta(A_{\sgrp},B_{\sgrp})$. This results in the partially
exchangeable prior on $\bbeta$ represented in Figure~\ref{AS}; the parameter $\nu_g$ allows sharing information on functionality across all variants in the same group.

As described in the appendix, the posterior conditional probability that a
variant $v$ belongs to the model depends on the overall number of groups, the number of groups considered relevant, and the number $\sumgrp_g=\sum_{\gamma(v)=g} \ivph_v$ of variants in the same group that are deemed functional.
%Let $\sumgrp_g=\sum_{\gamma(v)=g} \ivph_v$ denote the number of variants in
%group~$g$ that are associated with the trait.  The marginal posterior
%distribution is as in \eqref{eq:basicpost} except that the prior
%$f_{\svph}(\bvph)$ is replaced by the joint prior for $\bvph$ and $\bgrp$, which
%is given by \eqref{eq:priormulti} with the appropriate changes: $r$ in place of
%$p$, $p_g$ in place of $q$, and so forth. Group~$g$ partitions $\bvph$  into
%$\bvph_g$ (the elements of $\bvph$ in group~$g$) and $\bvph_{\mg}$.
%In a manner similar to the preceding approximations,
%$\expect[\igrp_g|\bgrp_{\mg},\bvph_{\mg},\tau,\by]^{-1}
%  \approx 1 + 1/h_g(\bgrp_{\mg},\tau,\by)$, where for $p_g>1$
%\begin{equation*}
%  h_g(\bgrp_{\mg},\tau,\by)
%    = \frac{A_{\sgrp}+\abs{\bgrp_{\mg}}}{B_{\sgrp}+r-\abs{\bgrp_{\mg}}-1}
%    \sum_{\bvph_g} [
%        \frac{\mathrm{B}(A_g+\abs{\bvph_g}, B_g+p_g-\abs{\bvph_g})}
%             {\mathrm{B}(A_g,B_g) (\tau\sqrt{n})^{\abs{\bvph_g}}}
%        \!\!\!\prod_{\grpsub} ( 1-\eta_{v}^2 )^{-n/2} ]
%\end{equation*}
%with the summation being over all $2^{p_g}$ possible values of $\bvph_g$.  When
%group~$g$ contains only one variant $v$, the summation is replaced by
%$( 1-\eta_{v}^2 )^{-n/2} / (\tau\sqrt{n})$.
The prior distribution in Figure~\ref{AS}, which we refer to as learning {\em
Across Sites}, allows one to achieve an effect similar to that of burden tests,
while still providing some variant-specific information (which is in contrast,
for example, to the proposal in \cite{YetL11}).

\section{Methods}
\label{sec:mcmc}

\subsection{MCMC sampling}

While we have resorted to some analytical approximation for expository
convenience,  we explore the posterior distribution with MCMC.
As previously mentioned,
we can focus on sampling  $\tau$ and all indicator variables. We use a Metropolis-within-Gibbs scheme, with the proposal distributions
described below.
For $\tau$,  the common practice of using a truncated Gaussian  works well.  The discrete indicator variables pose a
greater challenge, even though having integrated out $\bbeta$ allows us to work with sample space
of fixed dimension, eliminating the need for a reversible jump MCMC.  When
there  is only one layer of indicator variables $\bvph$, the proposal consists of first choosing with equal probability
whether to add or remove a variant and then choosing  uniformly among the
candidate variants the one for which to propose a change of status.  If the prior distribution is described using higher level
indicators as well, then proposed changes to both levels must be consistent. If
an entry of $\bvar$ is changed from one to zero, the associated entries of
$\bvph$ also have to be zeroed;  when proposing to change an entry of $\bvar$
from zero to one, the associated entries of $\bvph$ are selected from the
prior marginal.
Additionally, there are proposal moves that  leave $\bvar$ unchanged but then randomly
select one of its nonzero entries and draw a proposal for the associated entries
of $\bvph$ in a fashion analogous to that described previously.  Details of the
algorithm are in the supplementary material.

These simple proposal distributions will have trouble in two situations.  The
most common is when two or more variants are strongly associated with a
phenotype but are also strongly correlated with each other due to LD.  Any specific Markov chain
 will tend to include  one of the variants in the model, leaving out the rest.  Another problematic situation
is when the effects of two variants on a phenotype depend upon each other, so
neither variant is likely to enter the model by itself, even if their joint
inclusion would be favored by the posterior distribution.
Others \citep{GS2011,peltola1step,PeltolaAdapt} have described proposal
distributions that overcome these difficulties and that can be reasonably
applied to our setting---even though we do not 
investigate this in detail, focusing on the description of novel priors.

The average $\mbvph$ of realized
values of $\bvph$ can be used to summarize the evidence in favor of each variant. Given its practical importance,  the   basic convergence checks incorporated in our package are based on  $\mbvph$. By
default, the R code distributed in the package {\tt ptycho}
starts four chains from different points, runs each chain for a specified number
of MCMC iterations, computes the averages for each chain
separately, and then checks the range $\Delta\mbvph$ of these averages.
Details on the MCMC can be found in the supplementary material.
The algorithm is implemented in the R package {\tt ptycho} \citep{ptycho}.

\subsection{Evaluation of variant selection performance}

To investigate the performance of  the proposed priors, we apply them to simulated and real 
data. 
The posterior distribution can be summarized in multiple ways.
One can look for the indicator configuration that receives the highest
posterior, for example, or make marginal inference on each variant. Both
computational and robustness considerations make it practical to rely on
posterior averages $\mivph_{vt}$ for comparisons.
 In the Bayesian models, then, we consider selecting variant~$v$ for trait~$t$ if the  posterior average
$\mivph_{vt}$ is larger than a certain threshold $\xi \in (0,1)$:
$\mathcal{S}_t \equiv \{ v : \mivph_{vt} > \xi \}$.
For benchmarking purposes, we will also analyze the datasets with some
non-Bayesian approaches. Specifically we will consider (a) the Lasso
\citep{tibs1996}; (b) a set of univariate linear regressions (one for each
trait and variant), leading to $t$-statistics used to test the hypotheses
of no association $H_{vt}: \beta_{vt}=0$ with multiplicity adjustment for the
$pq$ hypotheses via the Benjamini-Hochberg (BH) procedure at level $\alpha$
\citep{BH};
and (c) multivariate regression including all possible variants, with
subsequent tests on the $pq$ null hypotheses for each coefficient incorporating
adjustment via the BH procedure at level $\alpha$.  The set of selected variants is
equivalent in (a)  to the set of estimated nonzero coefficients and in (b) and
(c) to the set of variants for which the $H_{vt}: \beta_{vt}=0$ are rejected. We
will refer to these approaches as (a) Lasso, (b) BH marginal, and (c) BH full. 

The threshold $\xi$ for Bayesian selection, the penalty of the Lasso, and the level $\alpha$ of BH can
all be considered tuning parameters.  We will compare the results of different
procedures as these are varied (see details in the supplementary material).  We
base our comparison on an empirical evaluation of power and FDR associated with the different methods. 
Specifically, for each simulation and each method of analysis, we calculate the
proportion of causal variants that are identified and the proportion of selected
variants that are in fact false discoveries. The average of these values across
multiple simulations is what we refer to as power and FDR in the results. The
Bayesian methods also provide an estimate of FDR: if $\mivph_{vt}$ is
approximately the probability that variant~$v$ is causal for trait~$t$, then
the mean of $(1-\mivph_{vt})$ over the selected variants is the Bayesian
False Discovery Rate.  We let $\hbfdr$ denote this mean and explore how well it
approximates (or not) the realized FDP, evaluated across all traits and
variants.

\subsection{Genotype and phenotype data}

Our work has been partially motivated by a resequencing study: 
\cite{SetF14} analyzed targeted exome resequencing data for 17 loci in
subjects of Finnish descent (from the 1966 Northern Finland Birth Cohort (NFBC)
and the Finland-United States Investigation of NIDDM Genetics study (FUSION)).
While the original study considered six quantitative metabolic traits, we focus
here on the fasting levels of  High Density Lipoprotein  (HDL), Low Density
Lipoprotein  (LDL), and triglycerides (TG), transformed and adjusted for
confounders as in the initial analyses (see supplementary material).  The
genotype data was obtained by sequencing the coding regions of 78~genes from
17~loci that had been found by previous GWAS meta-analyses to have a significant
association to one of the six traits.  In addition, we had access to  the
first five  principal components of genome-wide genotypes. The goal in \cite{SetF14} 
is to identify which variants in these loci are most likely to directly
influence the observed variability in the three distinct lipid traits.

Data cleansing and filtering are described in detail in the supplement; here we
limit ourselves to note that for the purpose of the simulation study,
the collection of variants was pruned to eliminate 550 variants observed only
once and to obtain a set of variants with maximal pairwise correlation equal to
0.3 by removing another 558 variants.  We excluded singletons from consideration
since it would not be possible to make inference on their effect without strong
assumptions. Multiple considerations motivated our choice of selecting a subset with
only  modest correlations: (a)~correlated variants make the convergence of MCMC
problematic, which might  impair our ability to understand the inference
derived from the posterior distribution; more importantly, (b)~it is very
difficult to evaluate and compare the performance of model selection methods in
the presence of a high correlation between variants; and finally, (c)~statistical
methods cannot really choose between highly correlated variants and the
selection among these needs to rely on experimental studies.
Let us expand on these last two points. Procedures
that build a multivariate linear model,  such as the Lasso, would select one out of
multiple highly correlated variants that have some explanatory power for the
response; approaches such as BH marginal would instead tend to select them all;
and Bayesian
posterior probabilities for each of the variants would reflect the fact that
substitutes are available: there will be multiple variants with elevated (if not very high in absolute terms) posterior probability.

It becomes difficult to meaningfully compare FDR and
power across these methods, substantially reflecting the fact that the problem
is somewhat ill-posed: if multiple highly correlated variants are available,
any of them can stand for the others, and it is arbitrary to decide on purely statistical grounds that one
belongs to the model while the others do not. Since our goal here is to
understand the operating characteristics of the procedures, we found it useful
to analyze them in a context where the target is well-identified and the results easily comparable.

After the described pruning, the genetic data used in the simulations contains
5335~subjects and 768~variants.  Genotypes were coded with minor allele counts,
and missing values (0.04\% of genotypes) were imputed using  variant average  counts for consistency with previous analysis. Observed minor allele
frequencies range from $2\times10^{-4}$ to 0.5, with a median of 0.0009 and a
mean of 0.02.  There are 628 variants with MAF<0.01.  Annotation information was
obtained as in \cite{SetF14}, resulting in 61\% coding, 34\% UTR, and the
remainder intragenic.  Prior to applying the selection methods, the five genetic principal
components along with the intercept were regressed out of both $\bX$ and $\bY$,
and the columns of both were then standardized.

When studying a real dataset, however, investigators might not be
comfortable with such a stringent level of pruning; one might be concerned that
variants with important effect are eliminated and that one is essentially reducing the information content of 
the sample. Indeed, when analyzing real data, we used a much more comprehensive approach, as described in the case study section.

\subsection{Simulation scenarios}

We constructed two simulation scenarios: one to simply 
illustrate the advantages of the proposed priors and the other to
 investigate their potential in a set-up that models a
real genetic investigation.

\subsubsection*{Illustrative example: orthogonal $\bX$.}

We set $n=5000$, $p=50$, $q=5$, and $\bX=\sqrt{\frac{n-1}{n/p}}
(I_p~I_p~\cdots~I_p)^T$ so that $\bX^T\bX=(n-1)I_p$.  In  generating $\bbeta$
and the responses, we want to cover a range of different signal-to-noise ratios.
To achieve this, we sample values of the parameters using the distributional
assumptions that we described in the specification of the priors. To explore the
performance of the {\em Across Traits} and {\em Across Sites} models---both when
they provide an accurate description of reality as well as when they do not---we
use three rules to generate the probability with which each variant is
associated to each trait. (a)~We sample one sparsity parameter $\beff$ for each
trait and keep it constant across variants. (b)~We sample a probability
$\meff_v$ for each variant and keep it constant across traits. Finally, (c)~we define
groups of five variants and sample one probability $\geff_g$ of causality for
each group of variants and each trait. Rules (a)--(c) are most closely reflected
in  the prior structure of the basic, {\em Across Traits} and {\em Across Sites}
models, respectively; and we indicate them as {\bf exchangeable variants}, {\bf
pleiotropy}, and {\bf gene effect}. We generate 100~datasets per rule, each with
$q$ responses, and analyze them with the described set of approaches. When using
Bayesian methods, we rely on non-informative priors (see supplement for
details).

\subsubsection*{Actual genotypes $\bX$.}

To explore the potential power and FDR in the analysis of the dataset with three
lipid traits, we generate artificial phenotypes starting from the available
pruned genotypes. We consider a mixture of possible genetic architectures. In
the construction of each dataset, (a) one gene is selected uniformly at random for
each phenotype and 3--4 of its rare variants are causal (gene effect); (b)  $40$
distinct common variants are selected uniformly at random and each has probability
equal to 0.1 to be causal for each of the phenotypes (thereby substantially
representing trait-specific variants); and, finally, (c) $10$ additional common variants
are selected uniformly at random and each has a probability 0.9 to be causal for
each phenotype (pleiotropic effects).  This results in traits that are on
average determined by 3--4 rare variants in one gene, 4 common variants with
effects on one trait only, and 9 common variants with effects across
multiple traits. We generated a total of 100 such datasets, as detailed in
the supplementary material.

%We also created another set of simulations along the lines of a traditional
%investigation of pleiotropy versus coincident linkage as described in the
%supplementary material.

\subsection{Data Availability}

The sequencing and phenotype data are available on dbGaP.  The Northern Finland
Birth Cohort 1966 (NFBC1966) study accession number is phs000276.v2.p1.  The
Finland-United States Investigation of NIDDM Genetics (FUSION) study accession
number is phs000867.v1.p1, with the sequencing data in substudy phs000702.v1.p1.
In both cases, the sequencing data used in this paper has molecular data type
equal to `Targeted Genome' rather than `Whole Exome'.

\section{Results}
\label{sec:results}

\subsection{Simulations}
\subsubsection*{Illustrative example.}
\begin{figure}[htbp]
\centering
\includegraphics[scale=0.8]{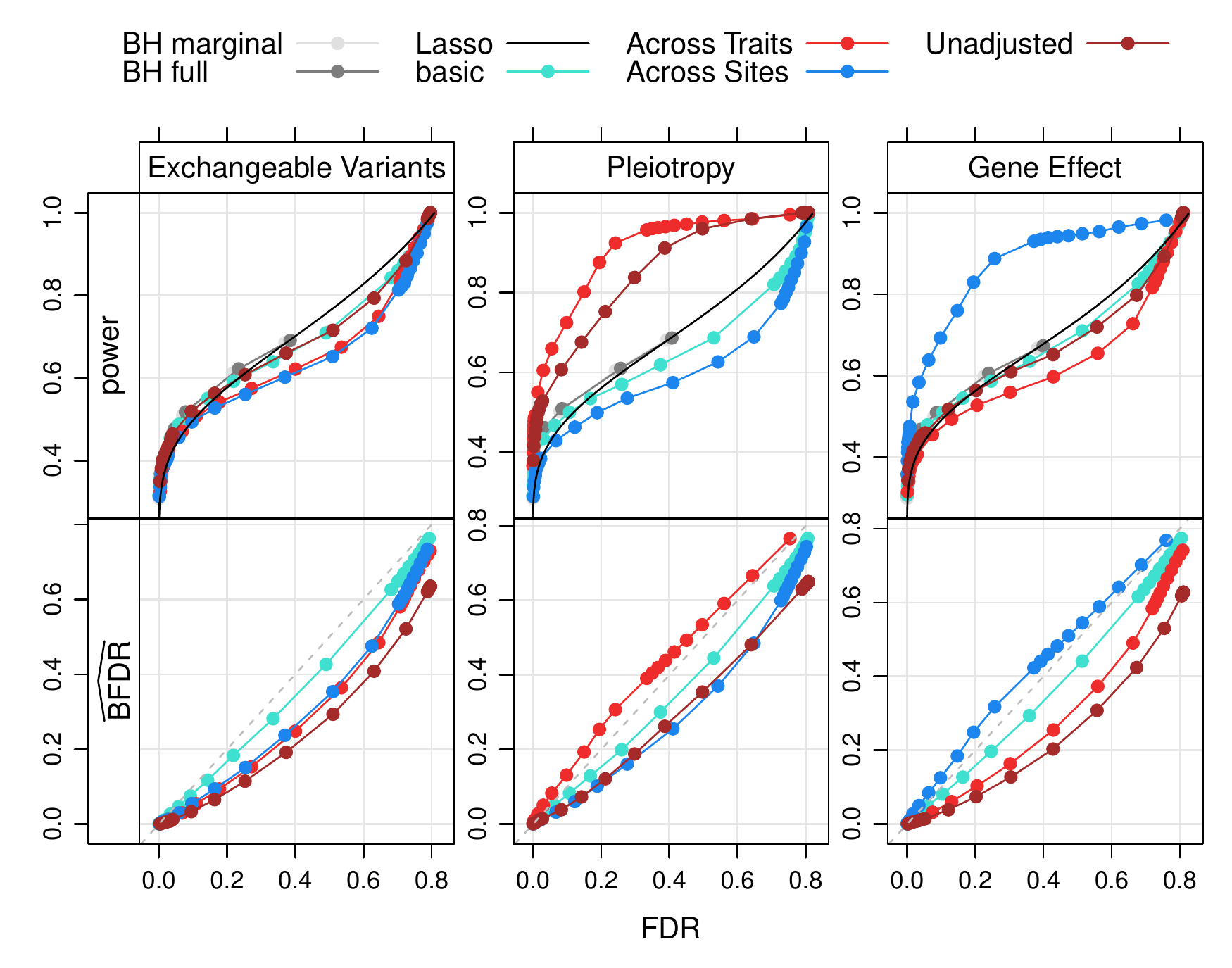}
\caption{\label{fig:bayesperf} Power (top) and $\hbfdr$ (bottom) as a function
of empirical FDR in the illustrative example.  Each color indicates a different
variant selection approach (see legend at the top). Displays in different
columns are for different data-generating mechanisms.}
\end{figure}

Figure~\ref{fig:bayesperf} showcases the possible advantages of the priors we
have described.
The plots on the top row compare the empirical FDR
and power of the different variant selection methods on the datasets with
orthogonal $\bX$. Points along the curves are obtained by varying tuning
parameters and averaging the resulting FDP and power across 100 simulated
datasets.  Our setting is such that BH full, BH marginal, Lasso and the basic
Bayes model have very similar behaviors:   the {\em Across Traits}
and {\em Unadjusted} models achieve the highest power per FDR in the presence
of pleiotropy and the worst power per FDR in the presence of gene effects;  in
contrast, the {\em Across Sites} model has maximal power in the
presence of gene effects and worse power in the presence of pleiotropy. While
it is not surprising that  the most effective prior is the one  that matches
more closely the structural characteristics of the data, it is of note that the loss of power
deriving from an incorrect choice of the {\em Across Traits} or the
{\em Across Sites} model is minimal for FDR values lower than 0.2,
which are arguably the range scientists might consider acceptable (see
supplementary information for a detail of these values).  On the bottom row of
Figure~\ref{fig:bayesperf}, we compare the estimated $\hbfdr$ with actual FDR
for the Bayesian models; here the most serious mistake is in underestimating
FDR, which would lead to an anti-conservative model selection. Once again it can
be seen that the best performance is obtained with the prior that matches
the data-generating process. Besides this, it is useful to
analyze the behavior of the {\em Unadjusted} approach: its power increase per
FDR in the presence of pleiotropy is less pronounced than that of the {\em
Across Traits} model, substantially because  the {\em Unadjusted}
approach is too liberal, with a $\hbfdr$ which is significantly underestimated.
This is in agreement with the lack of adjustment for multiplicity indicated by
\eqref{eq:multiEgammaj}.  Results for alternate hyperparameters are in the
supplement.

\subsubsection*{Generating phenotypes from actual genotype data.}

\begin{figure}[htbp]
\centering
\includegraphics[scale=0.8]{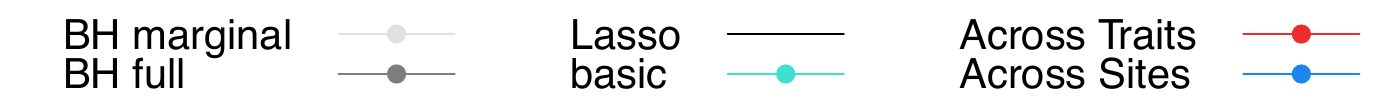}
\includegraphics[scale=0.8]{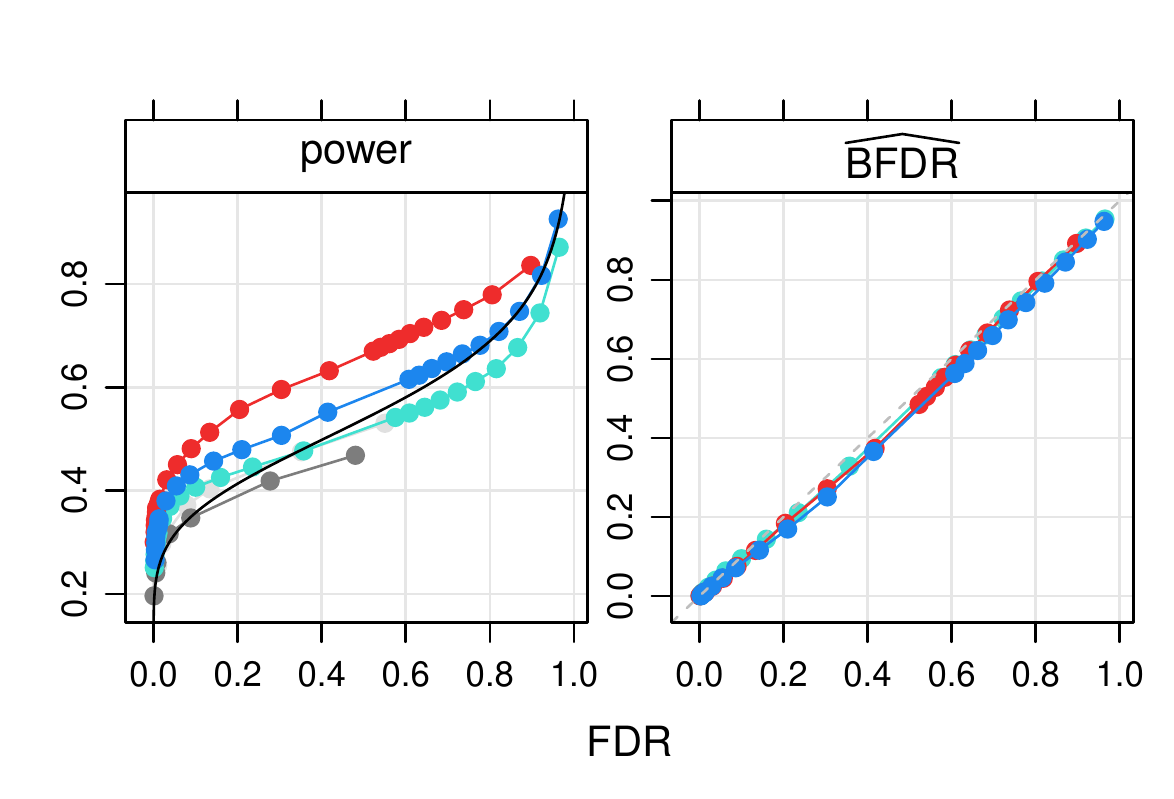}
\includegraphics[scale=0.8]{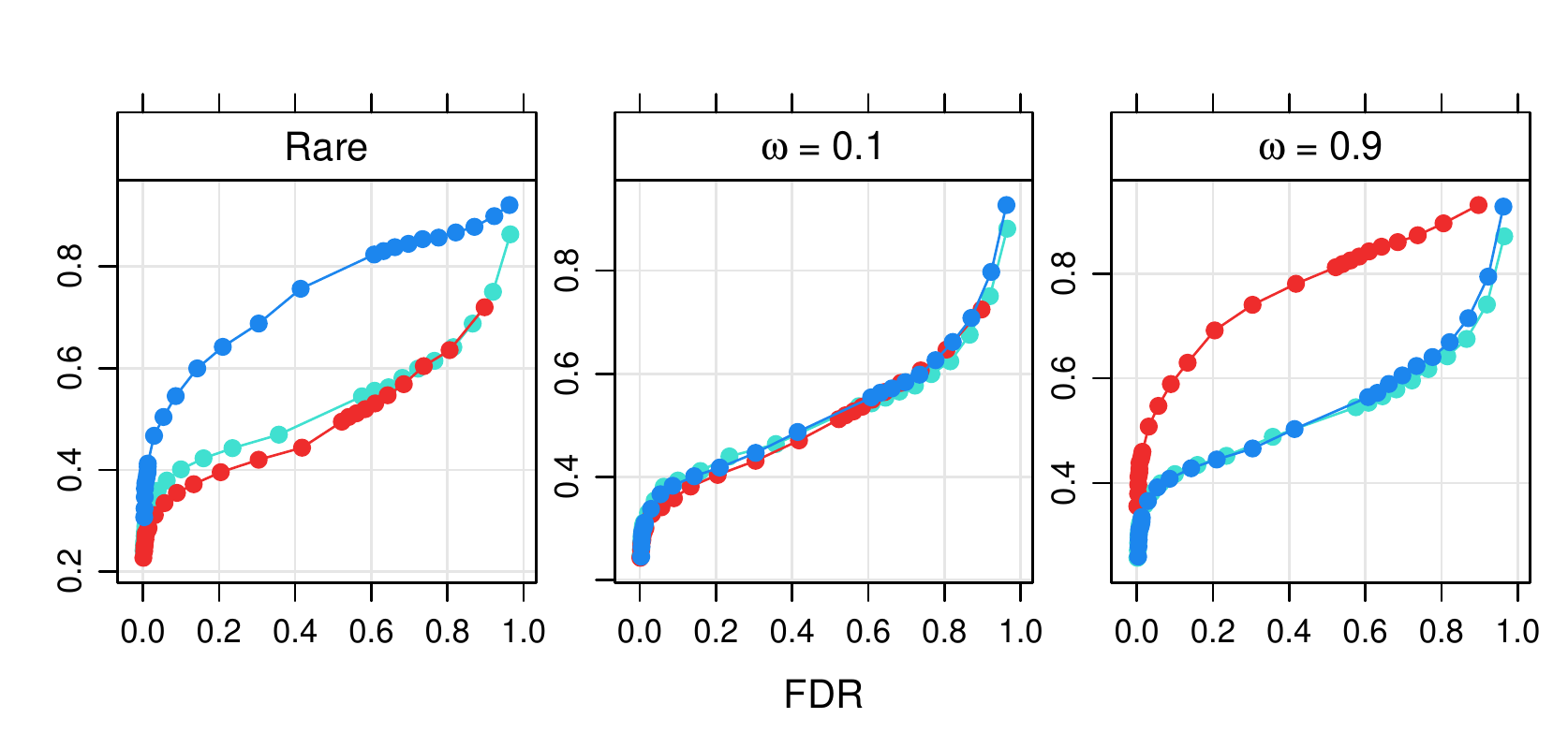}
\caption{\label{fig:actXperf} In the top portion, power and $\hbfdr$ as a
function of empirical FDR in the simulation from actual genotype data.  In the
lower panels, power is calculated separately for the different types of
variants: rare, common with trait-specific effects, and common with
pleiotropic effects.}
\end{figure}

Figure~\ref{fig:actXperf} shows the performance of the variant selection methods
in the analysis of traits generated from actual genotype data, further
emphasizing the potential gains associated with the proposed strategies.  
 For given FDR,
both the {\em Across Traits} and {\em Across Sites} priors lead to an
increase in power over the other methods.  This is due to the fact that
phenotypes are generated assuming both pleiotropy and contributions from multiple rare variants in the same gene (gene effects). In the
lower portion of Figure~\ref{fig:actXperf}, we separate the power to recover
rare variants with gene effects from that for trait-specific common variants
($\omega=0.1$) and from that for common variants with pleiotropic
effects ($\omega=0.9$).  As expected, the gains of {\em Across Traits} and
{\em Across Sites} are for the portion of genetic architecture that is
accurately reflected in these priors.  The estimates $\hbfdr$ are accurate,
indicating that all three Bayesian priors correctly learned $\tau$ and the
probabilities of function.

Finally, while we have relied on ROC-like curves to compare different approaches
as the value of their tuning parameters vary, it is useful to focus on the
operating characteristics of the standard ways of selecting the tuning parameters.
By convention, the target FDR for BH is usually 0.05.  For Lasso selection, the
function {\tt cv.glmnet} provides two choices for $\lambda$: minimizing the
cross-validation error and using the one-standard error rule.  In Bayesian
approaches, one can select variants so that global $\hbfdr \leq 0.05$.
Table~\ref{tbl:actXapriori} compares FDR and power for these selection
parameters; the Bayesian methods appear to control the target FDR and arguably
result in better power. Analogous summaries for other decision rules are
included in the supplementary material; here we simply remark that including
variants such that $\hbfdr \leq 0.05$ in this dataset was practically
equivalent to selecting variants with posterior probability larger than 0.7. We will capitalize on this observation for the real data analysis.
\begin{table}
\caption{\label{tbl:actXapriori} FDR and power for specific choices of selection
parameters applied to simulated traits with actual genotype data.}
\centering
\begin{tabular}{l|rr}
{\bf Variant selection criteria}  & {\bf FDR } & {\bf power }\\
\hline
BH full  with $\alpha=0.05$  & 0.037 & 0.32       \\ 
BH marginal with $\alpha=0.05$& 0.079 & 0.37       \\
\hline
Lasso min error $\lambda$& 0.810  & 0.63 \\
 Lasso 1-se $\lambda$ & 0.046 & 0.26\\
\hline
\vspace*{-.4cm} & &\\
basic           $\hbfdr \leq 0.05$       & 0.039 & 0.37\\
{\em Across Traits} $\hbfdr \leq 0.05$& 0.054 & 0.45\\
{\em Across Sites}   $\hbfdr \leq 0.05$  & 0.051 & 0.40
\end{tabular}
\end{table}

The supplement details the results of another set of simulations along the lines
of a traditional investigation of pleiotropy versus coincident linkage; we give
a very brief summary here.  In the case of separate causal variants, the {\em
Across Traits} prior may have a slight loss of power but is still much better
than BH with p-values from the full model.  In the case of pleiotropy, however,
the {\em Across Traits} prior clearly has greater power per FDR.

\subsection{Case Study: the influence of 17 genomic loci on lipid traits}
We now turn to the analysis of the three lipid traits in the Finnish dataset.
% In the following,
% as in \cite{SetF14}, variants  in dbSNP version~137 are indicated with their
% {\it rs} names, while other variants  are referred to as {\it
% v\_c9\_107548661}, specifying chromosome and position  (from GRCh37). 
While resequencing data comes from 17 loci identified via GWAS, prior
evidence of association is available only between some of these loci and some
traits. In particular, four loci have no documented association with any
of the three lipid traits we study; we include variants from these loci in the
analysis as negative controls.  (This is different from the work in
\cite{SetF14}, which examines only variants in loci specifically associated
with each trait.)

\cite{SetF14} relied on univariate regression to test the  association between
each trait and each variant with MAF>0.01 and on burden tests to evaluate
the role of nonsynonymous rare variants. \cite{BetC15} re-analyzed the data
relative to HDL with a set of model selection approaches, including the novel methodology SLOPE; to facilitate
comparison with their results, we add SLOPE to the analysis methods considered so
far. Groups for the {\em Across Sites} model were defined so as to mimic the
burden tests in \cite{SetF14}, which means a group with more than one variant
contains all nonsynonymous variants with MAF<0.01 in the same gene.

We start by analyzing the pruned subset of variants used in the simulation
studies and postpone a more exhaustive search, noting again that this
allows for a more straightforward comparison of the variants selected by
different methodologies.  Table~\ref{tbl:act3apriori} compares
the number of variants selected by various methods with specified tuning
parameters.  The column labeled $V^*$
shows the number of selected variants that are in a locus lacking any prior
evidence of association to lipid traits.
\begin{table}
\caption{\label{tbl:act3apriori}  Summary of selections for BH with
$\alpha=0.05$, Lasso with $\lambda$ chosen by {\tt cv.glmnet}, and Bayesian
approaches with $\xi=0.7$.
The columns labeled $R$ and $V^*$ give,
respectively, the  number of variants selected  across the entire study and in
the four loci with no prior evidence of association to any of the lipid
traits analyzed (CRY2, G6PC2, MTNR1B, and PANK1). $\hbfdr$ reports, for Bayesian
methods, the Bayesian FDR computed separately for each trait.}
\centering
\begin{tabular}{l||rrr|rrr|rrr}
& \multicolumn{3}{c|}{HDL} & \multicolumn{3}{c|}{LDL}
                                                    & \multicolumn{3}{c}{TG}\\
\hline
\hline
\vspace*{-.35cm}& & & & & & & & &
\\
 Variant Selection
 &  $R$&  $V^*$  &$\hbfdr$ & $R$&  $V^*$  &$\hbfdr$ &  $R$&  $V^*$  &$\hbfdr$\\
\hline  
BH full p-values    & 13 & 0 & &  3 &  0 & &  5 &  0 &\\
BH  marginal p-values  & 22& 0 & &  6 &  0 &  & 10 & 0 &\\
\hline
Lasso
min error $\lambda$ &134 &12 & & 40 & 4 & & 80 & 6& \\
Lasso 1-se $\lambda$ & 16 & 0 & &  0 & 0&  &  4 & 0&\\
\hline
basic                  & 21& 0 & 0.046 &  4& 0 & 0.012 &  8& 0 & 0.011\\
{\em Across Traits}& 19& 0 & 0.084 &  8& 0 & 0.059 &  9& 0 & 0.029\\
{\em Across Sites}     & 25& 0 & 0.064 &  5& 0 & 0.063 &  8& 0 & 0.007
\end{tabular}
\end{table}
 The Lasso with $\lambda$ chosen to minimize cross-validated prediction error
clearly results in far too many selections, so we discard this approach for the
remaining results.
For Bayesian approaches, the  threshold  $\xi=0.7$ results in average $\hbfdr$
approximately controlled at the $0.05$ level.
%(More extensive comparisons can be found in the supplementary material.)

\begin{figure}[htbp]
\centering
\includegraphics[scale=0.9]{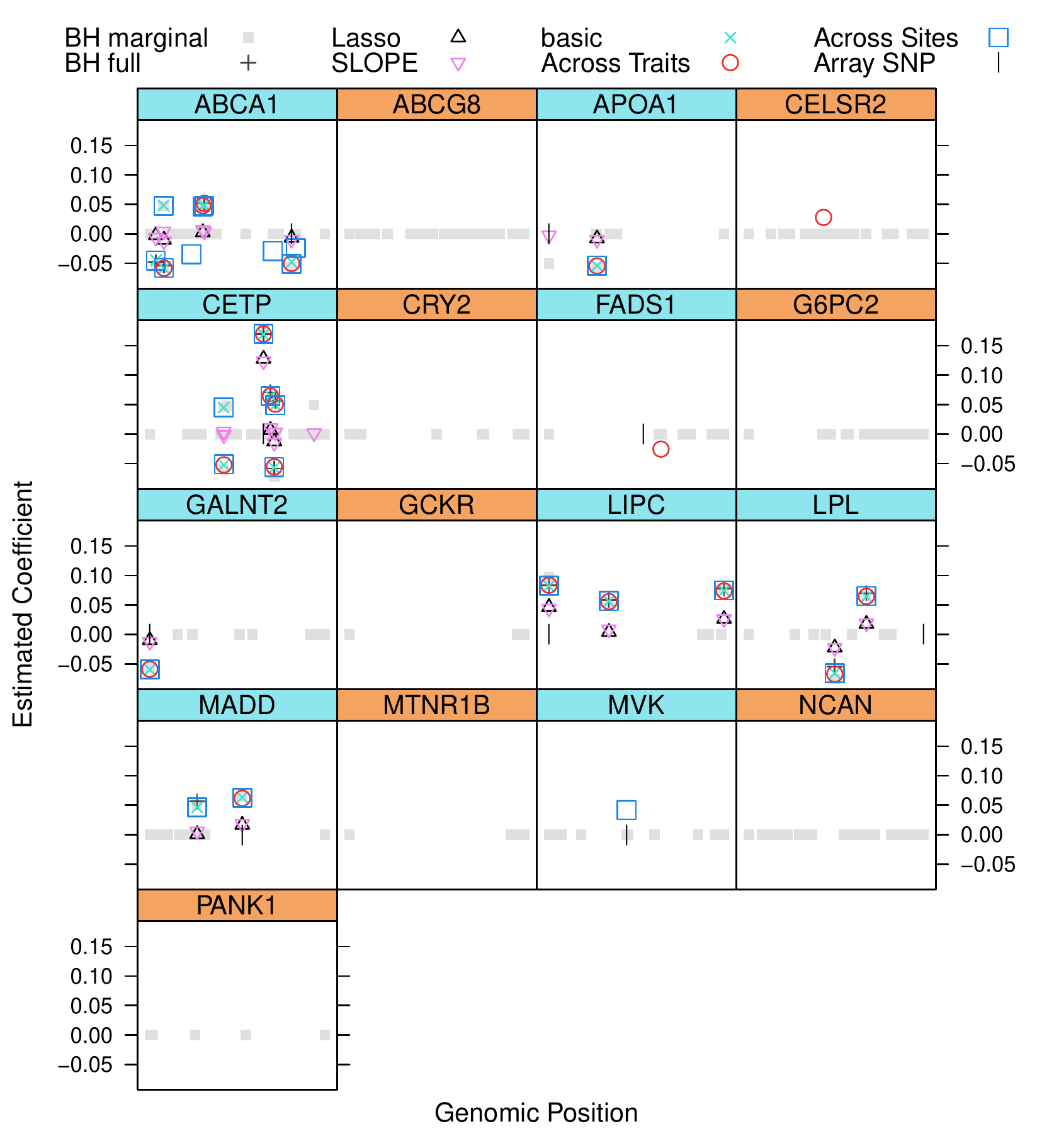}
\caption{\label{HDL} Estimated variant effects on HDL. Each panel corresponds
to a locus,  the $x$-axis indicates the variant's genomic position and the
$y$-axis its regression coefficient (with the exception of BH marginal, only
nonzero coefficients are represented). The color code of the panel titles
indicates the presence/absence of prior evidence of association between the
locus and HDL (turquoise/orange, respectively). Model selection methods are
distinguished using plotting symbols, as indicated in the legend at the top. }
\end{figure}
Figure~\ref{HDL} illustrates the model selection results for HDL.  (Analogous
displays for the other two phenotypes are available in the supplementary
material, as well as a table detailing differences in selections between approaches.) Each
display corresponds to a locus, with turquoise shading (rather than orange)
used to indicate  prior evidence of association to HDL. Variants are arranged
according to their genomic positions in the loci, and the values of their
estimated coefficients are plotted on the $y$-axis; with the exception of
marginal BH, we display only nonzero coefficients. When available, a vertical
black line indicates the position of the SNPs  originally used to select the
locus (`Array SNP').

There is substantial overlap among the results of various methods. Model
selection approaches seem to generally agree with the findings in
\cite{SetF14} (with Lasso 1-se the most discrepant, missing a number of the
associations identified in \cite{SetF14}; see supplementary material). Still, we
can point to some significant differences. With the {\em Across Traits}
approach we select two variants in two loci where no other method identifies
any signal: in CELSR2 and FADS1. These two loci have  prior evidence of
association to LDL and to all three lipid traits, respectively, and the {\em
Across Traits}
approach identifies pleiotropic effects.
In contrast, the {\em Across Traits} approach does not select four very rare
(MAF<0.001) variants considered relevant by more than one alternative method.
While we do not know where the truth lies at this point, it is very hard to
evaluate the effect of a rare variant on purely statistical grounds, and the
outcome of the  {\em Across Traits} model might well be the more reliable.  

The {\em Across Sites} approach identifies four variants that other approaches
overlook. Three are rare variants in ABCA1: two missense rare (0.01>MAF>0.001)
and one nonsense very rare (MAF 0.00016); their discovery is facilitated by the
fact that they are included in a group with multiple other significant variants.
The fourth is a common variant in the MVK locus, for which there is prior
evidence of association to HDL.  Other approaches do not recover this simply
because the signal in the locus is, in the dataset analyzed here, barely below
the detection threshold;  {\em Across Sites} has a slight advantage over the
other Bayesian methods because grouping reduces the number of comparisons to
account for. We note that ABCA1 is a gene in which rare variants were found to have a role by the burden test analysis in \cite{SetF14}.

We have relied on the pruned dataset since low correlation across variants
greatly facilitates the comparison of the selection results by different
methods. However, mindful of the concerns of scientists unwilling to pre-screen
the genotypic information, we have also carried out a more comprehensive
analysis of this dataset, showcasing that this is indeed an option available to researchers.  Details for these analyses are in the supplement, but
we summarize them here.  First, we have compared the results of four different
levels of pruning (correlation less than 0.3, 0.5, 0.7 and 0.9).  We have found
that very few values of $\mivph_{vt}$ change by more than 0.05 when different
levels of pruning are used and that less stringent levels of pruning do not lead
to substantially more findings---unlike when applying BH to marginal p-values.
In fact, there is a greater tendency for variants to drop out of the selection
set as correlated variants are added to $\bX$.  Second, to completely
eliminate pruning, we analyzed all variants in {\em one} locus  along the lines
of \cite{SerS07,HetE14,KetP14,CheL15}, using the basic prior and  assuming that
the number of significant regressors $\abs{\bvph}$ is no greater than five or
six (depending on the total number of possible variants in the locus).  We have restricted
our attention to two loci only, those that showed stronger evidence of
influencing HDL via multiple variants. The supplementary material offers a
precise comparison of results, but it suffices here to note that the set $\bvph$
of variants with the largest
posterior density is equal to the variants selected among the original pruned set (correlation less than 0.3) by the basic prior for one
locus and, for the other locus, the two sets substantially overlap.

\section{Discussion}

As the genetics community devotes increasing effort to follow up GWAS hits
with resequencing, a number of suggestions on how to prioritize variants
have emerged. In truth, while dealing with the same broad scientific goals, many
of these contributions address different aspects of the problem and therefore
should be seen as complementary rather than alternatives; taken together they
provide scientists with useful suggestions.  Annotation information has been
shown to be quite useful when the set of variants under consideration is
sufficiently diverse. It is important to account for the correlation across
variants to avoid paying attention to SNPs that are only `guilty by
association.' Bayes's theorem and Bayesian statistics are a natural way of
dealing with the decision of which variants to pursue. In this context,
others have studied (a)~how to choose priors that incorporate annotation
information, tuning their parameters with available datasets; (b)~how to
approximate posterior distributions of variant effects; (c)~how to sample from
the posterior distribution using efficient MCMC or variational schemes; and
(d)~how
to efficiently evaluate posterior probabilities for a set of variants.  Here we
focus on another aspect of prior selection: describing how partial
exchangeability assumptions can be used to borrow information across traits
and neighboring sites, while maintaining an effective control for multiplicity across
variants and fitting multivariate regression models that estimate the specific
contribution of each associated site, while accounting for others. We  briefly refer to some of the most
direct antecedents of our proposed priors to underscore relevant differences.

\cite{YetL11} proposed the use of hierarchical priors to capture effects of rare
variants through groups, similar to the {\em Across Sites} model.  However,
their proposal does not incorporate sparsity considerations, resulting in the
estimate of a nonzero effect for each variant and each group and therefore not
engaging in model selection.  \cite{QuiB11,QuiS12} took an additional step
towards the {\em Across Sites} model by incorporating sparsity via the indicator
variable $\bvph$.  They considered only rare variants, used the same effect
size for all rare variants in a genomic region, used the MLE for the effect
sizes rather than integrating them out, and, most importantly, controlled
sparsity by using $A_g=1$ and $B_g=p_g$ in the prior for $\meff_g$ rather than
introducing another layer of indicator variables in the hierarchical prior---all
of which means their approach has less flexibility and less learning.

The {\em Across Sites} prior also echoes the proposal of \cite{ZetL10} who
suggested the use of group penalization in Lasso to estimate multivariate sparse
models while encouraging coordinated selection of rare variants in the same
gene. This computationally appealing approach has not become as popular in
genetics as in many other fields, possibly because of the difficulties connected
with the selection of its tuning parameters when model selection is the goal.
Cross validation is often used to determine the appropriate level of
penalization; while this works well for prediction purposes, its performance is
less than satisfactory in identifying variants with truly nonzero coefficients
(as illustrated by our case study). \cite{AL11}, \cite{VetH12}, and,
most recently, \cite{SetV14} explore coupling resampling  techniques with Lasso
penalization to improve model selection. This not only increases computational
costs but also greatly reduces the initial simplicity of the model. 
% I changed here CS
As documented in \cite{BetC15}, identifying a single $\lambda$ value that
performs FDR control  is challenging;  \cite{YiB15} investigates this task in
the context of GWAS and provides guidelines. The final model selection of these
machine learning approaches uses complex rules; in contrast, the Bayesian
models we described are based on easy to interpret parameters.

% LLS + CS
The use of hierarchical Bayesian methods has ample precedents in eQTL studies,
where they have been used to correct for multiplicity \citep{KetA06} and
to increase power of detecting variants affecting multiple traits.  In our
presentation of the {\em Unadjusted} and {\em Across Traits} approaches, we
referred to methods proposed by  \cite{jia07} and \cite{bott11}.  More recent work
\citep{FetS13,nobel13} has focused on the identification of local (cis) effects
across tissue, and considered models with only one functional variant. 
The
recent contribution by \cite{CetZ14}---which appeared while this work was in
preparation---underscores as we do the importance of learning both across sites
and across traits to prioritize variants. These authors, however, work with
p-values from GWAS studies, rather than actual resequencing data. 

Having clarified the scope of our contribution, we want to briefly mention how it could be extended and combined with suggestions by others.
First, let us point out that while in the simulations and in the analytical
approximations  we assumed $n>p$, this restriction is by no means necessary to
the Bayesian model we describe. On  the contrary, the priors we propose---by
learning sparsity and giving positive probabilities to configurations with some
$\beta_v=0$---are well-suited to the case $n<p$. The real challenge in dealing with GWAS-type data would be from a computational standpoint: increased mixing for MCMC as described in
\cite{XetP14} and \cite{GS2011} or other algorithmic improvements
\citep{CS12,HetE14} would make our approach more widely applicable.

Another extension that is easily achieved is the combination of the {\em
Across Traits} and {\em Across Sites} priors. Most immediately, the group
indicators $\bgrp$ in Figure~\ref{AS} can be made trait-specific and linked
across phenotypes with the same approach used to link the $ \bvph_t$ in
Figure~\ref{AT}.

It is certainly possible to combine the partial exchangeability aspects of our
models with a prior that incorporates annotation information. Refer, for
example, to the {\em Across Traits} prior in Figure~\ref{AT}. Currently, the
distribution on $W_1,\ldots, W_p$, indicators of functionality of the variants,
is  a beta-binomial. However, it is trivial to change it to a mixture of
independent logits, with the linear model component including an intercept
effect---which would capture the overall sparsity---and a linear combination of
annotation indicators \citep{VetP08,KetP14,P14}.

Since our focus has been on specification of the prior, we have not paid
much attention to the data-generating model, which could certainly be improved.
Specifically, we want to underscore the fact that using a mixed-model approach
might be advisable to account for population structure \citep{KetE10} and when
analyzing many phenotypes whose quantitative value might be influenced by
confounders \citep{ZS14} or simply by genetic variants not included in the model.

In conclusion, we want to emphasize the increasing importance in human genetics
of models that account for pleiotropy. `Big data' in genetics has  often been
equated with the abundance of sequences, and these certainly pose a number of
management and interpretation challenges. Our increased acquisition capacity
will also result, however, in the collection of a large number of phenotypes;
gene expression, MRI scans, and mass spectrometry are just some examples of the
large-scale phenotyping efforts underway. Now that DNA variation has been
 extensively described, annotating this appears as a fundamental challenge;
the rich phenotypic collections increasingly available have  a major role to
play. After all, what better way of establishing if a variant has some functional
impact than looking for its association with any trait available?  Bayesian
models that allow one to estimate the probability with which a variant has
functional effects across phenotypes are likely to be very useful. In this
paper, we have described a first step in this direction.

\section*{Acknowledgements}
L. Stell and C. Sabatti  were partially supported by NIH grants HG006695,
HL113315, MH105578, and MH095454. We thank the authors of \cite{SetF14}
for letting us use their data during the completion of dbGaP release. Finally
our gratitude to H. Tombropoulos for unwavering editorial assistance.

\bibliographystyle{../arxiv/myplainnat}
\bibliography{bvswp}

\appendix

\section{Mathematical details for the \emph{basic} prior}

First we integrate $\bbeta$ out of the posterior distribution as in
\citep{CheL15}.  Since $\by=\bX_{\svph}\bbeta_{\svph}+\beps$ with
$(\bbeta_{\svph}|\bvph,\rho,\tau) \sim
\gauss\left(0,\frac{\tau^2}{\rho}\Sigma_{\svph}\right)$ and
$\beps \sim \gauss(0,\frac{1}{\rho}I_n)$,
\begin{equation}
  (\by|\bvph,\rho,\tau) \sim
     \gauss\left(0,
       \frac{1}{\rho}I_n
         + \frac{\tau^2}{\rho}\bX_{\svph}\Sigma_{\svph}\bX_{\svph}^T\right).
                        \label{eq:likelihood1}
\end{equation}
While the likelihood can be written directly from this, a few manipulations give
it in a more convenient form.
Sylvester's determinant theorem \citep[p.~420]{harville} implies
\begin{equation*}
  \det(I_n + \tau^2\bX_{\svph}\Sigma_{\svph}\bX_{\svph}^T)
    = \det(I_{\abs{\bvph}} + \tau^2\Sigma_{\svph}\bX_{\svph}^T\bX_{\svph})
    = \tau^{2\abs{\bvph}} \frac{\det(\Sigma_{\svph})}{\det(\Omega_{\svph})},
\end{equation*}
and a generalization of the Woodbury matrix identity \citep[p.~428]{harville}
implies
\begin{equation*}
  (I_n + \tau^2\bX_{\svph}\Sigma_{\svph}\bX_{\svph}^T)^{-1}
    = I_n - \bX_{\svph}
        (\tau^{-2}\Sigma_{\svph}^{-1} + \bX_{\svph}^T\bX_{\svph})^{-1}
              \bX_{\svph}^T
    = I_n - \bX_{\svph} \Omega_{\svph} \bX_{\svph}^T.
\end{equation*}
Consequently, $\pr(\by|\bvph,\rho,\tau) =
    \left(\frac{\rho}{2\pi}\right)^{n/2}
    \frac{\det(\Omega_\svph)^{1/2}}{\tau^{\abs{\bvph}}\det(\Sigma_\svph)^{1/2}}
           e^{-\rho S_\svph^2/2}$.
For the null model $\bvph=0$, \eqref{eq:likelihood1} shows that the covariance
matrix is $\rho^{-1}I_n$, so in this case $S_\svph^2 = \by^T \by$ and the ratio
of determinants is set equal to one.

Next multiply $\pr(\by|\bvph,\rho,\tau)$ by the prior density
function for $\rho$ and integrate to obtain
\begin{equation*}
     \frac{\det(\Omega_\svph)^{1/2}}
          {\tau^{\abs{\bvph}} \det(\Sigma_\svph)^{1/2}}
  \int \rho^{n/2}
       e^{-\rho S_\svph^2/2} \rho^{\alpha_\rho} e^{-\lambda_\rho \rho} d\rho
    \propto \frac{\det(\Omega_\svph)^{1/2}}
                 {\tau^{\abs{\bvph}} \det(\Sigma_\svph)^{1/2}}
        \left( \lambda_\rho + \frac{S_\svph^2}{2}
                    \right)^{-\left(\frac{n}{2}+\alpha_\rho\right)},
\end{equation*}
since the integrand is the density function of
$\tgamma\left( \alpha_\rho+\frac{n}{2},
\lambda_\rho + \frac{1}{2} S_\svph^2 \right)$, up to a normalizing factor.
Hence, the marginal posterior for $\bvph$ and $\tau$ is given by
\eqref{eq:basicpost}.

Along the lines of \cite{MalW11}, we present an approximation of the posterior
expected value $\expect[\ivph_v|\bvph_{\mv},\tau,\by]$.
If $\bvph$ and $\btvph$ are equal except that $\ivph_v=0$ and
$\tilde\ivph_v=1$ for one $v$, then
\begin{equation}
  \frac{f_{\svph,\tau}(\btvph,\tau|\by)}{f_{\svph,\tau}(\bvph,\tau|\by)}
    = \frac{\pr(\ivph_v=1|\bvph_{\mv},\tau,\by)}
           {\pr(\ivph_v=0|\bvph_{\mv},\tau,\by)}
    = \frac{\expect[\ivph_v|\bvph_{\mv},\tau,\by]}
           {1-\expect[\ivph_v|\bvph_{\mv},\tau,\by]}.
                       \label{eq:oddsratio}
\end{equation}
We will use several assumptions in order to simplify the expression on the left
and then solve for $\expect[\ivph_v|\bvph_{\mv},\tau,\by]$.
Consider the case when the columns of $\bX$
are orthogonal, which implies $\bX^T\bX\approx nI_p$ and the two choices of
$\Sigma_{\svph}$ are essentially the same.  Furthermore,
$\langle \bx_v, \by \rangle = \langle \bx_v, \bX \bbeta + \beps \rangle
    \approx n \beta_v + \langle \bx_v, \beps \rangle$,
which is distributed as $\gauss(0,\frac{n}{\rho} (\ivph_v n \tau^2 + 1) )$; in
this context, distinguishing signal from noise requires $n\tau^2 \gg 1$, so we
assume this to be the case.  Consequently,
$\Omega_{\svph}^{-1} \approx (n+\tau^{-2})I_p \approx nI_p$,
which in turn implies $S_\svph^2$ is approximately equal to
the residual sum of squares ($\rss$) for the model indicated by $\bvph$.
Finally, reflecting the results of current GWAS, we assume that the portion of variance explained (PVE) by the loci in consideration is rather small, so $\rss$ is not much less than $\by^T\by\approx n$ for
any model.  If one further chooses
$\alpha_\rho=\lambda_\rho \ll \frac{n}{2}$, then 
\begin{equation}
  \frac{f_{\svph,\tau}(\btvph,\tau|\by)}{f_{\svph,\tau}(\bvph,\tau|\by)}
    \approx \frac{1}{\tau}
            \frac{f_{\svph}(\btvph)}{f_{\svph}(\bvph)}
        \left(
            \frac{\det(\Omega_{\stvph}) \det(\Sigma_{\svph})}
                 {\det(\Sigma_{\stvph}) \det(\Omega_{\svph})} \right)^{1/2}
        \left( \frac{S_{\stvph}^2}{S_{\svph}^2} \right)^{-n/2}.
                                \label{eq:basicEgam1}
\end{equation}
Properties of the beta and gamma functions give that the ratio of $f_{\svph}$
values is $(A_\beff+\abs{\bvph}) / (B_\beff+p-\abs{\bvph}-1)$.  Furthermore,
$\sqrt{\det(\Omega_{\stvph}) \det(\Sigma_{\svph})\rule{0mm}{3mm}} /
    \sqrt{\det(\Sigma_{\stvph}) \det(\Omega_{\svph})\rule{0mm}{3mm}}
   \approx 1/\sqrt{n}$.
Finally, $S_\svph^2 \approx
  \by^T \by - \frac{1}{n} \sum_{u:\ivph_u=1} (\bx_u^T \by)^2$,
so
\begin{equation*}
  \frac{S_{\stvph}^2}{S_{\svph}^2}
    \approx \frac{S_{\svph}^2 - \frac{1}{n} (\bx_v^T \by)^2}{S_{\svph}^2}
    \approx  1 - \frac{(\bx_v^T\by)^2}{n\by^T\by} \equiv 1 - \eta_v^2.
\end{equation*}
Substituting these results into \eqref{eq:oddsratio} and \eqref{eq:basicEgam1}
gives \eqref{eq:basicEgammaj}.

\section{Mathematical details for learning across traits}

While the {\em Unadjusted} prior is not useful, we include its marginal
posterior density here for completeness.  Its derivation is very similar to that
of the basic model, so we focus on the differences.  A priori the rows of
$\bvph$ are independent and each has a beta-binomial distribution, so
$f_{\svph}(\bvph)
    = \prod_{v=1}^p \mathrm{B}(A_v+\sumvar_v, B_v+q-\sumvar_v) /
                         \mathrm{B}(A_v,B_v)$.
Furthermore, the columns of $\bY$ are independent given $\bvph$, $\bbeta$,
and $\rho$, and similarly for the columns of $\bbeta$; so
\begin{equation}
  f_{\svph,\tau}(\bvph,\tau|\bY)
    \propto f_\tau(\tau) f_{\svph}(\bvph)
        \prod_{t=1}^q \left( \lambda_\rho + \frac{S_{\bvph_t}^2}{2}
                     \right)^{-\left(\frac{n}{2}+\alpha_\rho\right)}
            \frac{\det(\Omega_{\bvph_t})^{1/2}}
                 {\tau^{\abs{\bvph_t}} \det(\Sigma_{\bvph_t})^{1/2}}.
                             \label{eq:postMultiNoZ}
\end{equation}
If $\bvph$ and $\btvph$ are equal except that $\ivph_{vt}=0$ and
$\tilde\ivph_{vt}=1$ for one $v$ and one $t$, then the same calculations as for
the basic model give that \eqref{eq:basicEgam1} simplifies as
\begin{equation*}
  \frac{f_{\svph,\tau}(\btvph,\tau|\by)}{f_{\svph,\tau}(\bvph,\tau|\by)}
    \approx \frac{1}{\tau\sqrt{n}} \frac{A_v+\sumvar_v}{B_v+q-\sumvar_v-1}
        (1 - \eta_{vt}^2)^{-n/2}.
\end{equation*}
This leads to the approximation \eqref{eq:multiEgammaj}.

Next we consider the {\em Across Traits} prior.  The posterior density
is the same as in \eqref{eq:postMultiNoZ} except that $f_{\svph}$ is replaced by
\begin{equation*}
\begin{split}
  f_{\svar,\svph}(\bvar,\bvph)
    &= \int \pr(\bvph|\bvar,\bmeff) \pr(\bvar|\vareff) \pr(\bmeff)
                          \pr(\vareff) d\bmeff d\vareff\\
    &= \frac{\mathrm{B}(A_\svar+\abs{\bvar},B_\svar+p-\abs{\bvar})}
            {\mathrm{B}(A_\svar,B_\svar)}
      \prod_{v:\ivar_v=1} \frac{\mathrm{B}(A_v+\sumvar_v,B_v+q-\sumvar_v)}
                           {\mathrm{B}(A_v,B_v)},     %\label{eq:priormulti}
\end{split}
\end{equation*}
provided that $\ivph_{vt}=0$ for all $v$ such that $\ivar_v=0$---otherwise,
$f_{\svar,\svph}(\bvar,\bvph)=0$.

To derive the approximation for the posterior expected values, consider $\bvar$
and $\btvar$ that are equal except that $\ivar_v=0$ and $\tilde{\ivar}_v=1$ for
one $v$.  Choose $\bvph$ consistent with $\bvar$, which means
$f_{\svar,\svph}(\bvar,\bvph)\neq0$.  Since we are using a subscript to denote a
column of a matrix, we use a superscript as in $\bvph^v$ to denote row $v$ of
$\bvph$.  Furthermore, $\bvph^{\mv}$ denotes the sub-matrix of $\bvph$ obtained
by deleting row $v$.  Straightforward modification of \eqref{eq:oddsratio} gives
\begin{equation*}
  \frac{\sum_{\btvph}f_{\svar,\svph,\tau}(\btvar,\btvph,\tau|\bY)}
       {f_{\svar,\svph,\tau}(\bvar,\bvph,\tau|\bY)}
    = \frac{\expect[\ivar_v|\bvar_{\mv},\bvph^{\mv},\tau,\bY]}
           {1-\expect[\ivar_v|\bvar_{\mv},\bvph^{\mv},\tau,\bY]},
                             %\label{eq:oddsratioZ}
\end{equation*}
where the summation is over all $\btvph$ such that $\btvph^{\mv}=\bvph^{\mv}$.
Furthermore, for any such $\btvph$,
\begin{equation*}
  \frac{f_{\svar,\svph,\tau}(\btvar,\btvph,\tau|\bY)}
       {f_{\svar,\svph,\tau}(\bvar,\bvph,\tau|\bY)}
    \approx \frac{A_\svar+|\bvar_{\mv}|}{B_\svar+p-|\bvar_{\mv}|-1}
      \frac{\mathrm{B}(A_v+\tilde{\sumvar}_v,B_v+q-\tilde{\sumvar}_v)}
           {\mathrm{B}(A_v,B_v)}
        \prod_{t:\tilde\ivph_{vt}=1} \frac{1}{\tau\sqrt{n}}
                  (1 - \eta_{vt}^2)^{-n/2}.
\end{equation*}
Hence,
$\expect[\ivar_v|\bvar_{\mv},\bvph^{\mv},\tau,\bY]^{-1}
    \approx 1 + 1/h_v(\bvar_{\mv},\tau,\bY)$ with
\begin{equation*}
 h_v(\bvar_{\mv},\tau,\bY)
    = \frac{A_\svar+|\bvar_{\mv}|}{B_\svar+p-|\bvar_{\mv}|-1}
   \sum_{\bvph^v} \left[
        \frac{\mathrm{B}(A_v+|\bvph^v|, B_v+q-|\bvph^v|)}
             {\mathrm{B}(A_v,B_v) (\tau \sqrt{n})^{|\bvph^v|}}
        \prod_{t:\ivph_{vt}=1} ( 1-\eta_{vt}^2 )^{-n/2} \right],
                 %\label{eq:multiZh}
\end{equation*}
where the summation is over all $2^q$ possible values of $\bvph^v$.

%Under additional assumptions, the expression for $h_v$ can be further
%simplified.  If $\eta_{jk}^2 \ll 1$, which seems very likely, then
%\begin{equation}
%  \prod_{k:\gamma_k^j=1} \left( 1-\eta_{jk}^2 \right)^{-n/2}
%    = \left( \prod_{k:\gamma_k^j=1} \left( 1-\eta_{jk}^2 \right) \right)^{-n/2}
%    \approx \left( 1 - \sum_{k:\gamma_k^j=1} \eta_{jk}^2 \right)^{-n/2}.
%\end{equation}
%Consequently, if $\sum_k \eta_{jk}^2$ is large enough, then the posterior
%expectation of $Z_j$ is close to one.
%As $\sum_{k=1}^q \eta_{jk}^2 \rightarrow 0$ and $q$ increases with $A_j=B_j=1$,
%the summation in the definition of $h_j$ will be dominated by the term with
%$\abs{\bgam^j}=0$, which implies
%\begin{equation}
%  h_j(\bZ_{\mj},\tau,\bY)
%    \sim \frac{A_0+\abs{\bZ_{\mj}}}{B_0+p-\abs{\bZ_{\mj}}-1}
%            \frac{\mathrm{B}(1, 1+q)}{\mathrm{B}(1,1)}
%    = \frac{1}{q+1} \frac{A_0+\abs{\bZ_{\mj}}}{B_0+p-\abs{\bZ_{\mj}}-1}.
%\end{equation}

\section{Mathematical details for learning across sites}

The derivation of the marginal posterior for learning {\em Across Sites}
consists of straightforward modifications of previous calculations.  The
marginal posterior distribution is as in \eqref{eq:basicpost} except that the
prior $f_{\svph}(\bvph)$ is replaced by the joint prior for $\bvph$ and $\bgrp$,
which is
\begin{equation*}
  f_{\sgrp,\svph}(\bgrp,\bvph)
    = \frac{\mathrm{B}(A_\sgrp+\abs{\bgrp},B_\sgrp+r-\abs{\bgrp})}
            {\mathrm{B}(A_\sgrp,B_\sgrp)}
      \prod_{\genfrac{}{}{0pt}{}{g:\igrp_g=1,}{p_g>1}}
               \frac{\mathrm{B}(A_g+\sumgrp_g,B_g+p_g-\sumgrp_g)}
                    {\mathrm{B}(A_g,B_g)}
\end{equation*}
provided that $\ivph_v=0$ if $\igrp_{\gamma(v)}=0$, and
$\ivph_v=1$ if $\igrp_{\gamma(v)}=1$ and $p_{\gamma(v)}=1$---otherwise,
$f_{\sgrp,\svph}(\bgrp,\bvph)=0$.  Similar to the preceding approximations,
$\expect[\igrp_g|\bgrp_{\mg},\bvph_{\mg},\tau,\by]^{-1}
\approx 1 + 1/h_g(\bgrp_{\mg},\tau,\by)$, where for $p_g>1$
\begin{equation*}
  h_g(\bgrp_{\mg},\tau,\by)
    = \frac{A_\sgrp+\abs{\bgrp_{\mg}}}{B_\sgrp+r-\abs{\bgrp_{\mg}}-1}
    \sum_{\bvph_g} \left[
        \frac{\mathrm{B}(A_g+\abs{\bvph_g}, B_g+p_g-\abs{\bvph_g})}
             {\mathrm{B}(A_g,B_g) (\tau\sqrt{n})^{\abs{\bvph_g}}}
        \prod_{\grpsub} ( 1-\eta_{v}^2 )^{-n/2} \right]
\end{equation*}
with the summation being over all $2^{p_g}$ possible values of $\bvph_g$.  When
$p_g=1$, however, the summation is replaced by
$( 1-\eta_{v}^2 )^{-n/2} / (\tau\sqrt{n})$, where $v$ is the lone variant in
group~$g$.  Hence, the posterior conditional probability of $\sgrp_g$
depends upon the overall number of groups and the number of groups considered
relevant, while the posterior conditional probability of $\svph_v$ given
that $\sgrp_g=1$ depends on the number $\sumgrp_g$ of variants in the same group
that are deemed functional.

\newpage
\centerline{\bf Supplementary Material}

%\vspace*{-1em}
\noindent
\underline{\bf Choice of hyperparameters}
%\vspace{0.5em}

We state here the hyperparameters used in most of our computations; the effect of
alternate hyperparameters is discussed below.  For fine-mapping, there is
considerable uncertainty about the likely value of $\beff$, so we use the
uninformative prior with $A_\beff=B_\beff=1$.
Since $n\geq5000$ in our examples, we use $\alpha_\rho=\lambda_\rho=10$ to
ensure that $\alpha_\rho = \lambda_\rho \ll \frac{n}{2}$.

The bounds on the uniform prior for $\tau$ justify additional explanation.  The
lower bound could be zero, but this may lead to increased computational times
because very small $\tau$ implies small nonzero values in $\bbeta$ are more
likely, which may make the MCMC iteration accept larger models.  The assumption
$n\tau^2\gg1$ suggests choosing $\tau_1 \approx 1/\sqrt{n}$ instead.
Furthermore, $\tau^2/\rho$ is approximately the variance of the nonzero
coefficients since the diagonal entries of either $\Sigma_\svph$ are
approximately one.  The standardization of $\by$ and the assumption of small PVE
implies $\rho\approx1$.  Consequently, $\tau_2$ should be at least three times
as large as the largest likely value of $\abs{\beta_v}$.  Either marginal or
multivariate linear regression of our actual data gives all coefficients with
absolute value less than 0.2, so we use $\tau_1=0.01$ and $\tau_2=10$ in our
analyses.

%\vspace{3em}
\noindent
\underline{\bf MCMC sampling}
%\vspace{1em}

We use Gibbs sampling to alternate between updating $\tau$ and updating the
indicator variables.  We use Metropolis-Hastings for each update.  We only
update the indicator variables once for each update of $\tau$, but it might be
more efficient to try multiple indicator variable jumps between each jump of
$\tau$.

To define the notation, the Metropolis-Hastings algorithm to generate samples
with density proportional to $f(\cdot)$ draws a proposal $\tilde{u}$
from a jumping distribution $Q(\tilde{u}|u^{(m)})$, where $u^{(m)}$ is the
previous sample.  Next, compute
\begin{equation}
  R = \frac{Q\left(u^{(m)}\big|\tilde{u}\right)}
           {Q\left(\tilde{u}\big|u^{(m)}\right)\rule{0mm}{3mm}}
      \frac{f\left(\tilde{u}\right)}{f\left(u^{(m)}\right)\rule{0mm}{3mm}}.
                    \label{eq:mh}
\end{equation}
The new sample is $\tilde{u}$ with probability $\min(R,1)$ and is $u^{(m)}$
otherwise.  To avoid overflow, we actually compute $\log{R}$ and then take its
exponential.  The remainder of this section gives the definition of the jumping
distributions used.

The jumping distribution for $\tilde\tau$ given $\tau$ is
$\gauss(\tau,(\tau_2-\tau_1)^2/16)$ but truncated to the interval
$(\tau_1,\tau_2)$.  When $\abs{\bvph}=0$, the marginal posterior density is
independent of $\tau$ in $(\tau_1,\tau_2)$.  Hence, if all effect sizes are very
small, the $\tau$ samples may be essentially uniform.  This is because the data
do not enable learning the effect sizes and has nothing to do with the MCMC
algorithm or its convergence.
%Consequently, the variance of the $\tau$ samples should be checked before using
%them for inference.

With one level of indicator variables, our proposal distribution
$Q(\btvph|\bvph)$ is nonzero only if $\btvph-\bvph=\delta\be_v$ for
$\delta\in\{-1,+1\}$ and one $v$, where $\be_v$ is the vector with entry~$v$
equal to one and all others equal to zero.  Furthermore, for such $\bvph$ and
$\btvph$,
\begin{equation}
  Q_\ivph(\btvph|\bvph) = \left\{ \begin{array}{ll}
                  \frac{1}{p} & \text{if } \abs{\bvph}=0
                                \text{ or } \abs{\bvph}=p\\
                  \frac{\pi_+}{p-\abs{\bvph}}
                     & \text{if } \delta=+1\\
                  \frac{1-\pi_+}{\abs{\bvph}}
                     & \text{if } \delta=-1\\
                        \end{array} \right.,    \label{eq:mhjump}
\end{equation}
where $\pi_+$ is the probability of adding a variable to the model.

Next we define $Q_{\ivar,\ivph}(\btvar,\btvph|\bvar,\bvph)$ for learning \emph{Across Traits}.
The first step at each iteration is to choose $\delta$ from $\{-1,0,+1\}$ with
probabilities $\pi_-^*$, $\pi_0^*$, and $\pi_+^*$, respectively, with obvious
modifications if, say, $\delta=-1$ is not possible because $\abs{\bvar}=0$.
Then $\btvar-\bvar = \delta \be_v$, where variant~$v$ is chosen uniformly from
the possible candidates.  The proposal $\btvph$ will be the same as $\bvph$
except for changes to $\bvph^v$.  If $\delta=-1$, $\btvph^v=\bzero$.  If
$\delta=0$, \eqref{eq:mhjump} gives $Q_{\ivph}^*(\btvph^v|\bvph^v)$ with $q$
replacing $p$.  If $\delta=+1$, then draw $\btvph^v$ based on the prior for
$(\bvph|\bvar)$: first draw $\sumvar_v$ from the beta-binomial distribution
\begin{equation}
  \pr(\sumvar_v)
    = \frac{\mathrm{B}(A_v+\sumvar_v, B_v+q-\sumvar_v)}{\mathrm{B}(A_v,B_v)}
                 \binom{q}{\sumvar_v}            \label{eq:mhbetabin}
\end{equation}
and then sample $\sumvar_v$ distinct entries uniformly from $\{1,\ldots,q\}$.
The overall jumping probability is $Q_{\ivar,\ivph}(\btvar,\btvph|\bvar,\bvph)
 = Q_0(\btvph|\btvar,\bvar,\bvph)\, Q_{\ivar}(\btvar|\bvar)$, where
\begin{equation}
  Q_0(\btvph|\btvar,\bvar,\bvph) = \left\{ \begin{array}{ll}
                  1 & \text{if } \delta=-1\\
                  Q_{\ivph}^*(\btvph^v|\bvph^v) & \text{if } \delta=0\\
             \frac{\mathrm{B}(A_v+\sumvar_v, B_v+q-\sumvar_v)\rule{0mm}{2.6mm}}
                       {\mathrm{B}(A_v,B_v)\rule{0mm}{2.6mm}}
                     & \text{if } \delta=+1
                        \end{array} \right.    \label{eq:mhjumpsub}
\end{equation}
and the nonzero values of the mass function for $(\btvar|\bvar)$ are
\begin{equation}
  Q_{\ivar}(\btvar|\bvar) = \left\{ \begin{array}{ll}
                  \hat\pi_W & \text{if } \delta = 0\\
                  \frac{1}{p} & \text{if } \abs{\bvar}=0\\
                  \frac{\pi_-^*\rule{0mm}{2.6mm}}{(1-\pi_+^*) p\rule{0mm}{2.8mm}}
                            & \text{if } \abs{\bvar}=p \text{ and } \delta=-1\\
                  \frac{\pi_+^*\rule{0mm}{2.6mm}}{p-\abs{\bvar}\rule{0mm}{2.8mm}}
                          & \text{if } \delta=+1 \text{ and } \abs{\bvar}\neq0\\
                  \frac{\pi_-^*\rule{0mm}{2.6mm}}{\abs{\bvar}\rule{0mm}{2.8mm}}
                          & \text{if } \delta=-1 \text{ and } \abs{\bvar}\neq p
                        \end{array} \right..    \label{eq:mhjumpZ}
\end{equation}
The value of $\hat\pi_W$ is irrelevant because it will be the same for
$Q_W(\btvar|\bvar)$ and $Q_W(\bvar|\btvar)$, which will consequently cancel each
other in \eqref{eq:mh}.  Furthermore, if $\delta=+1$, then
$Q_0(\btvph|\btvar,\bvar,\bvph)$ in the denominator in \eqref{eq:mh} will cancel
the leftover factor in $f_{\ivar,\ivph}(\btvar,\btvph)$ in the numerator;
whereas for $\delta=-1$ a factor in $f_{\ivar,\ivph}(\bvar,\bvph)$ cancels
$Q_0(\bvph|\bvar,\btvar,\btvph)$.

The mass functions for the \emph{Across Sites} prior are more complicated, but
the fortunate cancellations still occur.  First, $p_g$ replaces $p$ in
\eqref{eq:mhjump} to give $Q_{\ivph}^*(\btvph_g|\bvph_g)$ when $\delta=0$, and
\eqref{eq:mhjumpsub} becomes
\begin{equation}
  Q_0(\btvph|\btgrp,\bgrp,\bvph) = \left\{ \begin{array}{ll}
                  1 & \text{if } \delta=-1\\
                  Q_{\ivph}^*(\btvph_g|\bvph_g) & \text{if } \delta=0\\
                  1 & \text{if } \delta=+1 \text{ and } p_g=1\\
             \frac{\mathrm{B}(A_g+\sumvar_g, B_g+p_g-\sumvar_g)\rule{0mm}{2.6mm}}
                       {\mathrm{B}(A_g,B_g)\rule{0mm}{2.6mm}}
                     & \text{if } \delta=+1 \text{ and } p_g > 1
                        \end{array} \right..
\end{equation}
Let $\igrp_g^*=1$ if $\igrp_g=1$ and $p_g>1$; otherwise it is zero.  Assuming that
at least one group has $p_g>1$, the nonzero values of the mass function for
$(\btgrp|\bgrp)$ are
\begin{equation}
  Q_{\igrp}(\btgrp|\bgrp) = \left\{ \begin{array}{ll}
                  \hat\pi_G & \text{if } \delta = 0\\
                  \frac{1}{r} & \text{if } \abs{\bgrp}=0\\
                 \frac{\pi_-^*\rule{0mm}{2.6mm}}{(1-\pi_+^*) r\rule{0mm}{2.8mm}}
                            & \text{if } \abs{\bgrp}=r \text{ and } \delta=-1\\
                 \frac{\pi_+^*\rule{0mm}{2.6mm}}{r-\abs{\bgrp}\rule{0mm}{2.8mm}}
                        & \text{if } \delta=+1 \text{ and } \abs{\bgrp^*}\neq0\\
                 \frac{\pi_-^*\rule{0mm}{2.6mm}}{\abs{\bgrp}\rule{0mm}{2.8mm}}
                          & \text{if } \delta=-1, \abs{\bgrp^*}\neq 0,
                             \text{ and } \abs{\bgrp}\neq r\\
                 \frac{\pi_+^*\rule{0mm}{2.6mm}}{(1-\pi_0^*)(r-\abs{\bgrp})\rule{0mm}{2.8mm}}
                        & \text{if } \delta=+1, \abs{\bgrp^*}=0
                             \text{ and } \abs{\bgrp}\neq 0\\
                 \frac{\pi_-^*\rule{0mm}{2.6mm}}{(1-\pi_0^*)\abs{\bgrp}\rule{0mm}{2.8mm}}
                          & \text{if } \delta=-1, \abs{\bgrp^*}=0,
                             \text{ and } \abs{\bgrp}\neq r
                        \end{array} \right..
\end{equation}

In our experiments, $\pi_+=0.5$ for the $\bvph$ proposal, while $\pi_0^*=0.5$ and
$\pi_-^*=\pi_+^*=0.25$ for the $\bvar$ or $\bgrp$ proposal.  One chain starts with
no variants in the model while another starts with all variants in the model.
For the other two chains, the variants are ranked by the absolute value of
their correlation with $\by_t$, and the $J_t$ most strongly correlated have
$\ivph_{vt}$ set to one.  For the actual phenotypic data, $J_t$ is 10 for one chain
and 20 for the other.  For the simulations, one chain has $J_t$ equal to the
expected number of variants that are causal for trait~$t$ (as determined by the
$\beff$ or $\bmeff$ used to generate $\bvph_t$), and the other chain has $J_t$
double that.  The burn-in interval was 10,000 iterations.  For the simulated
data, 500,000 samples were then used to compute the averages, which took a
couple of hours or less (typically much less, especially when using only one
trait at a time) running the chains sequentially.  For the actual data, 10 times
as many samples were used for the averages; the chains were run in parallel and
each MCMC sampling took 2--3~hr.  We always used the g-prior.

To assess convergence of the averages $\mivph_{vt}$, we computed the average for
each chain and then set $\Delta\mivph_{vt}$ equal to the difference between the
maximum and minimum averages over the four chains.  For each prior applied to
each set of 100 datasets, at least 95\% of $\Delta\mivph_{vt}$ were less than
0.05.  For the priors that incorporate $\bvar$ or $\bgrp$, at least 95\% of the
analogous values of $\Delta\overline{\ivar}_v$ or $\Delta\overline{\igrp}_g$
were less than 0.05 except that only 92\% of $\Delta\overline{\igrp}_g$
satisfied this condition when the {\em Across Sites} prior was applied to the
traits simulated from the genotype data.

%\vspace{3em}
\noindent
\underline{\bf Evaluation of variable selection performance}
%\vspace{1em}

The p-values for use in the BH procedure are from ordinary least squares, using
either one variant at a time (`marginal') or all variants simultaneously
(`full').  For each dataset, we apply BH to all traits simultaneously,
which means each BH test has $pq$ hypotheses.  For the performance curves, we
choose a target FDR, say $\alpha$, and apply BH with that $\alpha$ to each
dataset, computing the false discovery proportion (FDP) and of discoveries of
true null hypotheses for each dataset.  We then average over the 100
datasets to obtain the FDR and power for one point in our plots.  For our
performance curves, we used the FDR control targets $\{
0.001,0.005,0.01,0.05,0.1,0.3,0.5 \}$.  Applying BH to the traits separately
gave essentially the same results for the simulated data; the effect of testing
traits separately in the actual data is discussed below.

The performance curves for the Lasso show the effect of varying the penalty
parameter $\lambda$.  We use the R function {\tt glmnet} with its default
parameters (with {\tt family='gaussian'}) except without an intercept.
This returns the fits for a set of $\lambda$ values, but that set depends upon
the input data.  We compute the FDP and power at each value of $\lambda$ output
for that trait.  Using {\tt locpoly()} in the R package {\tt KernSmooth}, we
perform local linear regression on all these points for all datasets to
approximate FDP and power as functions of $\lambda$.  When instead choosing a
specific $\lambda$ by cross-validation, the one-standard error rule gives the
largest value of $\lambda$ such that the cross-validation error is within one
standard error of the minimum error.  In this case, the choice depends upon the
randomly drawn cross-validation bins, so for reproducibility we set the random
seed to 1234 immediately before calling {\tt cv.glmnet}.  

To compute the performance curves for Bayesian selection of variants, we choose
threshold $\xi$, select using the rule $\overline\ivph_{vt} > \xi$ for each
dataset, compute the FDP and power for each dataset (same as for BH), and
then average over all 100 datasets to obtain one point in our plots.  The
values of $\xi$ were $0.01 \ell$, $1-0.01\ell$, and $0.1 \ell$ for
$\ell=1,\ldots,9$.

%\vspace{3em}
\noindent
\underline{\bf Genotype and phenotype data}
%\vspace{1em}

Variants in dbSNP version~137 are indicated with their {\it rs} names, while
other variants  are referred to as {\it v\_c9\_107548661}, say, specifying
chromosome and position (from GRCh37). 
Starting with the genotype and phenotype data in \cite{SetF14},
we removed 786 individuals that were missing values for any of the principal
components or the three traits HDL, LDL, and log-transformed TG, leaving 5335
subjects.  From the sequencing data that had passed quality control, we removed
two variants that were missing data for over 20\% of the subjects as well as
variants whose minor allele occurred only once in the remaining subjects,
leaving 1326 variants.

Additional filtering involved: (1) annotation, whose possible values were
nonsense, missense, coding synonymous, UTR~$5^\prime$, UTR~$3^\prime$, intron,
and unknown; (2) PolyPhen2 predictions, whose possible values were probably
damaging, possibly damaging, benign, and unknown; (3) marginal regression
p-values from \cite{SetF14}; and (4) minor allele frequency (MAF).
If all subjects have exactly the same genotype (including missing values) for
two or more variants, we chose one of the variants using the ad hoc rules in
Figure~\ref{fig:ident}, thereby eliminating 24 more variants.  This $\bX$ was
used in the exact computation of the posterior probabilities in individual loci
discussed briefly at the end of the Results section of the main paper.

For the MCMC method of estimating the posterior probabilities across all loci
simultaneously, we further pruned to obtain a set of variants with maximal
correlation less than a specified bound $\cmax$, for reasons discussed in the
main paper.  Figure~\ref{fig:filter} gives our algorithm to achieve this.  The
set $U_0$ contains variants mentioned in \cite{SetF14} and \cite{BetC15}.  It
was introduced so that we only dropped such variants when they were correlated
with each other, which simplified comparison with those papers.  Furthermore, we
manually chose which of two variants to drop in some cases, as we now explain.
\cite{SetF14} list 39 associations between an `Array SNP' and a trait.  Of
those, 16 had p-value less than 0.001 for their data and involved one of the
three traits HDL, LDL or TG.  Our filtering process dropped one of these:
{\it rs10096633}, which was an Array SNP for both HDL and TG, because it had
correlation 0.96 with {\it rs328}, which has a nonsense mutation.  Among the new
discoveries reported by \cite{SetF14}, {\it rs651821} and {\it rs2266788} in
gene APOA5 are both associated with one of our traits, but we had to drop
one of them because their correlation is 0.95.  We dropped the latter because it
had a slightly stronger correlation with the Array SNP {\it rs12805061}.

\begin{figure}[p]
\center
\framebox{%
\footnotesize
\vbox{\begin{tabbing} 
\hspace{2em}\= \hspace{2em}\= \hspace{2em}\= \hspace{2em}\= \kill
For each of the 23 sets of variants for which all subjects have the same
genotype\\
\> If they are on the same chromosome and have the same annotation and PolyPhen2
prediction\\
\>\> Choose one arbitrarily\\
\> Else if exactly one variant has annotation `missense' (none had `nonsense')\\
\>\> Choose it\\
\> Else if no variant has annotation `missense'\\
\>\> Choose the one with annotation `coding synonymous' (which meant discarding
UTR~$3^\prime$ variant or\\
\>\> variant without annotation)\\
\> Else (this case only occurred once)\\
\>\> Choose the one with PolyPhen2 prediction of `probably damaging' (which
meant discarding the variant\\
\>\> predicted to be benign)
\end{tabbing}}%
}
\caption{\label{fig:ident} Pseudocode for eliminating duplicate variants.}
\end{figure}

\begin{figure}[p]
\center
\framebox{%
\footnotesize
\vbox{\begin{tabbing} 
\hspace{2em}\= \hspace{2em}\= \hspace{2em}\= \hspace{2em}\= \kill
For mode = 1,2\\
\> For $c = 0.9, 0.8, \ldots, \cmax$\\
\>\> variants = var.filter(variants, mode, $c$)\\
\\
\\
{\bf Function} var.filter$(V,m,c_0)$\\
$C$ is the absolute value of the correlation matrix of $V$, computed with the R
option {\tt use="pairwise.complete.obs"}\\
Consider the entries of $C_{ij}$ in decreasing order\\
\> If $C_{ij} \leq c_0$, return $V$\\
\> $V^* = \{ v_i, v_j \}$\\
\> If (m == 1)\\
\>\> If $v_i$ and $v_j$ are on different chromosomes, continue\\
\>\> Add to $V^*$ all $v_\ell$ on same chromosome as $v_i$ and such that
$C_{i\ell}$ or $C_{j\ell}$ is greater than $c_0$\\
\> $U_0$ contains any entries in $V^*$ that we particularly want to keep (see
text)\\
\> $U_1 = \argmin_{v_i\in V^*} \text{ p-value}$---but only consider common
variants\\
\> $U_2$ contains variants in $V^*$ with the `worst' annotation, where the
ordering is nonsense, then missense with\\
\> probably damaging, then missense with possibly damaging, then missense with
any prediction\\
\> If $U_0$ is not empty\\
\>\> $U$ is the entry in $U_0$ with the highest priority\\
\> Else if $U_1$ is empty\\
\>\> $U=U_2$\\
\> Else if $U_2$ is empty\\
\>\> $U=U_1$\\
\> Else\\
\>\> $U=U_1 \cap U_2$\\
\>\> If $U$ is empty\\
\>\>\> $U=U_1$\\
\> If $U$ is empty, $U=V^*$\\
\> Stop in the following rules when one gives single variant $u$:\\
\>\> 1. $u \in U$\\
\>\> 2. $u$ is variant in $U$ with greatest MAF\\
\>\> 3. Choose $u$ from $U$ arbitrarily\\
\> Drop $u$ from $V$ and $C$
\end{tabbing}}%
}
\caption{\label{fig:filter} Pseudocode for filtering correlated variants.}
\end{figure}

For almost all of our results, $\cmax=0.3$.  In this case, after filtering, we
put {\it rs12805061} and {\it rs3135506} (which is a missense mutation predicted
to be probably damaging) back into $\bX$, which then had 768 variants.  After
imputation, all pairwise correlations in $\bX$ are less than 0.3 except for two
that are slightly greater.  There are 628 variants with MAF\textless0.01.
To explore the effects of pruning the variants, we also consider $\bX$
obtained by setting $\cmax$ to 0.5, 0.7 and 0.9, resulting in 968, 1042 and 1124
variants, respectively.

Using the phenotype data (via the p-values) when filtering the variants could
affect the accuracy of methods to estimate or control FDR, but our previous
experience has been that this does not occur in this situation.

Because substituting means for missing genotype values is not accepted practice,
we comment on our use of this approach.  Since we discard the two
variants with very many missing values, only 0.04\% of the values in $\bX$
(with $\cmax=0.3$) are missing, and only six variants are missing data for more
than 1\% of subjects---one variant is missing data for 2.2\% of subjects and the
next worst is missing 1.3\%.  With $\cmax=0.9$, only 0.08\% of the values in
$\bX$ are missing.

Each of the traits (or its logarithm) were regressed on age, age$^2$, and
indicator variables for sex, oral contraceptive use, pregnancy status, and
cohort.  The five genetic principal components along with the intercept were
regressed out of both $\bX$ and the residuals $\bY$ from this regression,
and the columns of both were then standardized.

%\vspace{3em}
\noindent
\underline{\bf Simulation scenarios}
%\vspace{1em}

For the datasets with orthogonal $\bX$, we set $n=5000$, $p=50$, $q=5$, and
$\bX=\sqrt{\frac{n-1}{n/p}} (I_p~I_p~\cdots~I_p)^T$ so that $\bX^T\bX=(n-1)I_p$.
The columns of $\bX$ were not centered because that would destroy the
orthogonality.  In generating phenotypic data, $\rho=1$, $\tau \sim
\tunif(0.045,0.063)$, nonzero $\beta_{vt}\sim\gauss(0,\tau^2)$, and the
distributions of the probabilities of association were as follows:
\begin{description}
\item[Exchangeable variants] one $\beff \sim \tbeta(12,48)$ was drawn
independently for each trait.
\item[Pleiotropy] one $\vareff \sim \tbeta(16,55)$ was drawn for each
dataset, $\ivar_v$ was drawn for each variant, and then $\meff_v \sim
\tbeta(48,12)$ was drawn for each nonzero $\ivar_v$.
\item[Gene effect] $\grpeff \sim \tbeta(16,55)$ was drawn independently for
each trait, $\igrp_g$ was drawn for each group of five consecutive
variants, and then $\geff_g \sim \tbeta(48,12)$ was drawn for each nonzero
$\igrp_g$.
\end{description}
These choices ensured $n\tau^2\gg1$ so that signal was greater than noise,
$\expect[\ivph_{vt}]\approx0.2$, and $\abs{\bvph_t}$ exceeded 15 only about
5--15\% of the time.  Very roughly
\begin{equation*}
  \langle \by_t, \by_t \rangle
    = \langle \bX_{\bvph_t} \bbeta_{\bvph_t},
                   \bX_{\bvph_t} \bbeta_{\bvph_t} \rangle
          + 2 \langle \bX_{\bvph_t} \bbeta_{\bvph_t}, \beps_t \rangle
          + \langle \beps_t, \beps_t \rangle
    \approx n \langle \bbeta_{\bvph_t}, \bbeta_{\bvph_t} \rangle
          + \langle \beps_t, \beps_t \rangle
    \approx \frac{n}{\rho} \left( \tau^2 \abs{\bvph_t} + 1 \right),
\end{equation*}
so the generated traits also had variance close to one---between 0.94 and
1.17.  There were 100 datasets, each with $q=5$ traits.  When applying the
{\em Across Sites} prior to any of these data sets, the groups are the same as
when generating the dataset with gene effect: the first group consists of the
first five variants, the second group consists of the next five variants, and so
forth.

When simulating phenotypes from the actual genotype data, all rare
(MAF\textless0.01) variants in the same gene with missense or nonsense mutations
are in the same group.  Every other variant---common or rare---is in a group by
itself.  Then $\bvph$ was generated randomly as follows:
\begin{itemize}
\item for each trait~$t$, draw one group from amongst those with at least
five variants and set $\ivph_{vt}=1$ for 3--4 rare variants (number drawn
uniformly) in that group
\item draw 10 common variants and, for every trait~$t$, set $\ivph_{vt}=1$
with probability 0.9
\item draw 40 common variants and, for every trait~$t$, set $\ivph_{vt}=1$
with probability 0.1
\end{itemize}
The rest of the process to generate each trait was the same as in the other
simulations.  Again, there were 100 datasets, each with $q=3$ traits.
There are 20 genes with at least five nonsynonymous rare variants; 16 of them
have fewer than 10 variants, but the others have 14--16 variants.

%LAUREL TO ADD DESCRIPTION OF PLEIOTROPY-LINKAGE EXPERIMENT.
We also created another set of simulations along the lines of a traditional
investigation of pleiotropy versus coincident linkage, which we will call the
pleiotropy-linkage experiment.  We began with $\bX$ created by setting
$\cmax=0.7$ in Figure~\ref{fig:filter}.  We then took pairwise correlations
between all variants with MAF \textgreater0.1.  From the 10 pairs with absolute
correlation greater than 0.6 and in the same locus, we selected the five pairs
listed in Table~\ref{tbl:pairs}.  For each selected pair, we simulated 40
datasets, each with $\bX$ restricted to the one locus and with two traits.  In
half the datasets, one of the variants is causal for both traits
\begin{table}
\caption{\label{tbl:pairs} Pairs of variants used in the pleiotropy-linkage
experiment.  The total number of variants in the datasets for that pair is $p$.}
\centering
\begin{tabular}{llll|cr|r}
\multicolumn{4}{c|}{variants (MAF)} & locus & $p$ & correlation\\
\hline
rs11142    & (0.29) & rs464218 & (0.45) & CELSR2& 128 & 0.70\\
rs567243   & (0.49) & rs560887 & (0.31) & G6PC2 &  62 & -0.66\\
rs15285    & (0.24) & rs316    & (0.12) & LPL   &  25 & 0.63\\
rs12445698 & (0.19) & rs5801   & (0.18) & CETP  &  88 & 0.65\\
rs1801706  & (0.21) & rs5882   & (0.39) & CETP  &  88 & 0.66
\end{tabular}
\end{table}
(pleiotropy, with each variant causal equally often) while the other datasets
have one variant causal for one phenotype and the other variant causal for the
other phenotype (coincident linkage).  No other variants were causal.  The rest
of the process to generate each trait was the same as in the other simulations.

%\vspace{3em}
\noindent
\underline{\bf Results for illustrative example}
%\vspace{1em}

\begin{figure}
\begin{center}
\includegraphics[scale=0.8]{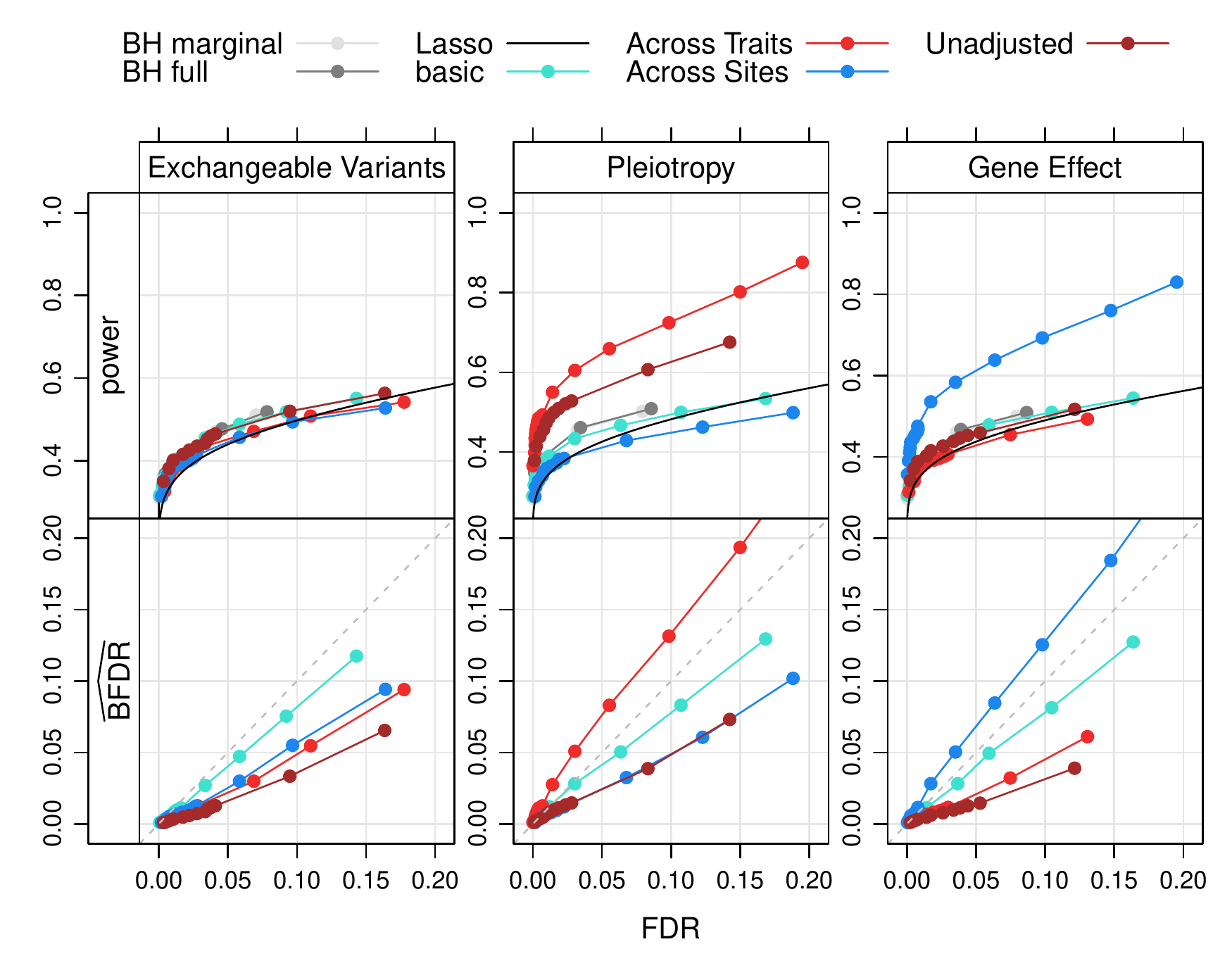}
\end{center}
\caption{\label{fig:bayeszoom} Power (top) and $\hbfdr$ (bottom) as a function
of empirical FDR in the illustrative example, limiting to FDR\textless0.2.  Each
color indicates a different variable selection approach (see legend at the top).
The different columns correspond to different data generating mechanisms.}
\end{figure}

For our Bayesian variable selection methods and several non-Bayesian methods
applied to the datasets with orthogonal $\bX$, Figure~\ref{fig:bayeszoom}
compares the power for FDR less than 0.2.  In this range, the power of the
{\em Across Sites} prior applied to the dataset with pleiotropy is only
slightly less than that of the basic prior or the non-Bayesian methods.
Similarly, the loss of power of the {\em Across Traits} prior applied to the
dataset exhibiting gene effects is slight when FDR is less than 0.2.

\begin{figure}
\begin{center}
\includegraphics[scale=0.8]{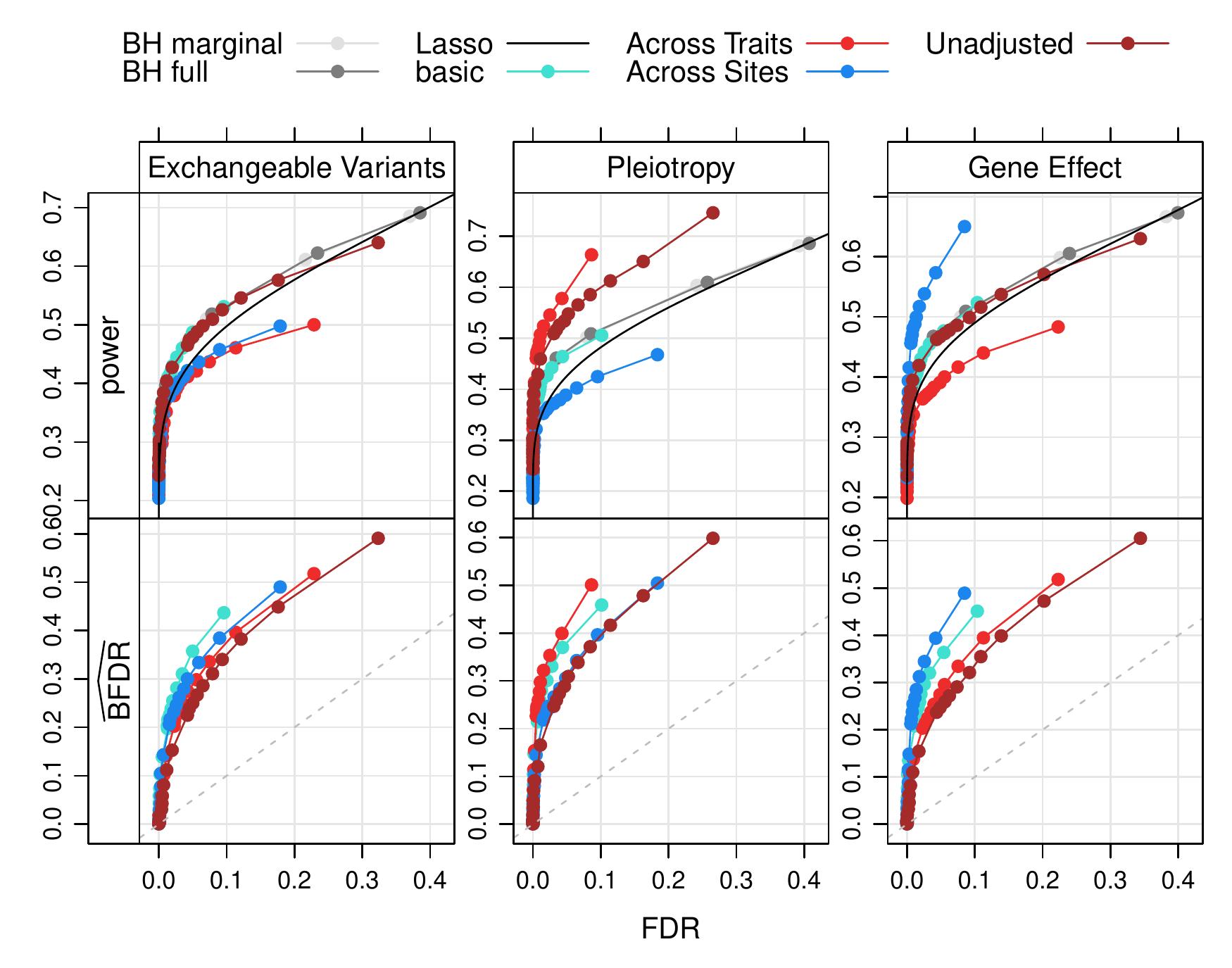}
\end{center}
\caption{\label{fig:tau} Power (top) and $\hbfdr$ (bottom) as a function
of empirical FDR in the illustrative example with prior $\tau\sim\tunif(2,5)$.
Each color indicates a different variable selection approach (see legend at the
top).  The different columns correspond to different data generating
mechanisms.}
\end{figure}

We also investigated the effect of using other hyperparameters.
Figure~\ref{fig:tau} shows the results when the prior for $\tau$ was
$\tunif(2,5)$ but everything else was the same as for the results in
Figure~\ref{fig:bayeszoom}.  As suggested by the approximations given in the
main paper for the conditional posterior expected values of the indicator
variables, increasing $\tau$ causes
$\overline\ivph_{vt}$ to decrease; so fewer variables are selected for given
$\xi$, decreasing both FDR and power.  This also means that
$\overline\ivph_{vt}$ underestimates the probability that variant~$v$ is causal
for trait~$t$, and $\hbfdr$ greatly overestimates FDR.  On the other hand, using
the prior $\tau\sim\tunif(0.045,0.063)$ as was used in data generation gave
essentially the same results as in Figure~\ref{fig:bayeszoom} because the mean and variance
of the $\tau$ samples were usually less than 0.1 when using the uninformative
prior $\tau\sim\tunif(0.01,10)$.

\begin{figure}
\begin{center}
\includegraphics[scale=0.8]{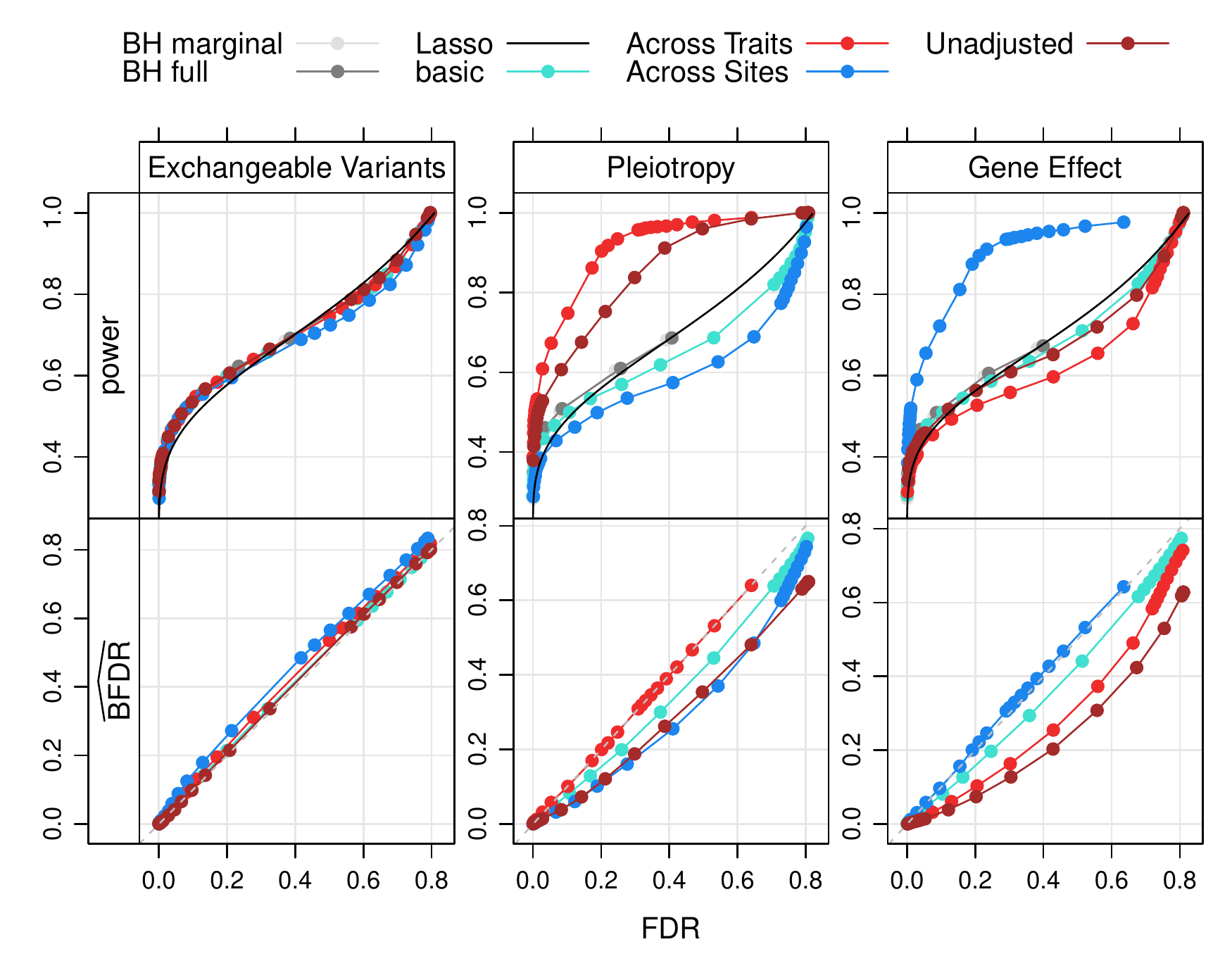}
\end{center}
\caption{\label{fig:omprior} Power (top) and $\hbfdr$ (bottom) as a function
of empirical FDR in the illustrative example with tighter priors for the
probability of causality in some cases.  Each color indicates a different
variable selection approach (see legend at the top).  The different columns
correspond to different data generating mechanisms.}
\end{figure}

Finally, we tried using different hyperparameters for the priors on the
probabilities of causality but with everything else the same as for the results
in Figure~\ref{fig:bayeszoom}.  For the dataset with exchangeable variants, we again used
uninformative priors for $\vareff$ and $\grpeff$ but all other Beta priors
had parameters $A=12$ and $B=48$.  For the dataset with
pleiotropy, we changed only the hyperparameters for $\beff_{\svar}$ and
$\bmeff$ for the
\emph{Across Traits} prior, which now have the same values as during data
generation, and likewise for the dataset with gene effects and the \emph{Across
Sites} prior.  Figure~\ref{fig:omprior} shows that this did not have much effect
on the results.  This suggests there is not much advantage to using tighter
priors, even if it were possible to guess the hyperparameters so accurately.

%\vspace{3em}
\noindent
\underline{\bf Results for generating phenotypes from actual genotype data}
%\vspace{1em}

When applying the {\em Across Sites} prior to data with actual genotype data,
the groups were the same as described above for simulating phenotypes from this
data.  This mimicked the burden tests in \cite{SetF14}.

To create the performance curves in the lower row of plots in Figure~5 in the
main paper, the FDR for each $\xi$ is equal
to that computed for the curves for all variants, but the power is computed
separately for each type of variant.
Table~\ref{tbl:actXapriori2} shows the FDR and power when using specified values
of $\xi$ to select variants after applying the Bayesian priors to the same set
of simulations.  The results when $\xi=0.7$ are essentially the same as the
results in the paper from controlling $\hbfdr\leq0.05$.

\begin{table}
\caption{\label{tbl:actXapriori2} FDR and power for selected values of $\xi$ when
applying Bayesian methods to simulated phenotypes with actual genotype data.}
\centering
\begin{tabular}{l||rr|rr|rr}
           & FDR  & power & FDR  & power & FDR  & power\\
\hline
\hline
      & \multicolumn{2}{c|}{$\xi=0.5$} & \multicolumn{2}{c|}{$\xi=0.7$}
                      & \multicolumn{2}{c}{$\xi=0.8$}\\
\cline{2-7}
basic                  & 0.100 & 0.41 & 0.039 & 0.37 & 0.021 & 0.35\\
{\em Across Traits}& 0.133 & 0.51 & 0.057 & 0.45 & 0.031 & 0.42\\
{\em Across Sites}     & 0.143 & 0.46 & 0.054 & 0.41 & 0.029 & 0.38
\end{tabular}
\end{table}

\begin{figure}
\begin{center}
\includegraphics[scale=0.7]{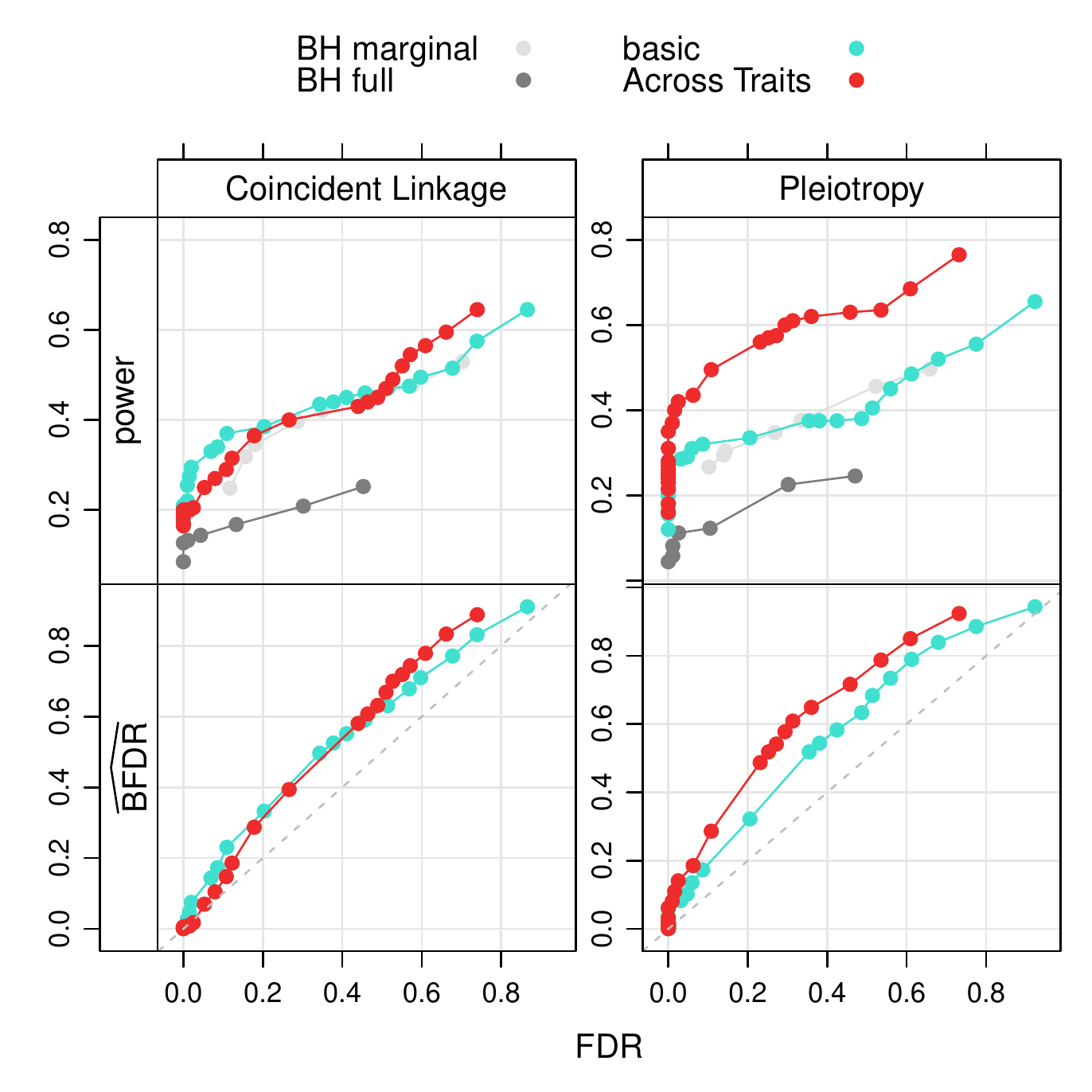}
\end{center}
\caption{\label{fig:pleiotropy} Power (top) and $\hbfdr$ (bottom) as a function
of empirical FDR in the experiment to check whether {\em Across Traits} can
distinguish between pleiotropy (right column) and traits with different
correlated causal variants (left column).  Each color indicates a different
variable selection approach (see legend at the top).}
\end{figure}

%LAUREL TO ADD DESCRIPTION OF PLEIOTROPY-LINKAGE EXPERIMENT.
Next we discuss the results of the pleiotropy-linkage experiment.  Over all the
datasets, at least 95\% of $\Delta\mivph_{vt}$ were less than 0.05 after 500,000
MCMC iterations, and when using the {\em Across Traits} prior at least 90\% of
$\Delta\overline{\ivar}_v$ were.  Figure~\ref{fig:pleiotropy} shows the FDR and
power for the basic and {\em Across Traits} priors as well as BH applied to both
sets of p-values.
In the case of separate causal variants, the {\em Across Traits} prior may have
a slight loss of power when FDR is less than 0.5 but is still much better than
BH with p-values from the full model.  In the case of pleiotropic traits,
however, the {\em Across Traits} prior clearly has greater power per FDR than
the other methods, which have roughly the same power regardless of whether the
traits have the same or different causal variants.  For the variants other than
the pair listed in Table~\ref{tbl:pairs}, 8\% of the time the difference in
$\mivph_{vt}$ between the two priors is less than 0.05.  When the traits are
pleiotropic, the differences are similar for the non-causal variant; but for the
causal variant, $\mivph_{vt}$ for the non-causal variant is sometimes greater
than 0.7 with the {\em Across Traits} prior even when it is less than 0.2 with
the basic prior, consequently increasing power considerably.  When the traits
are not pleiotropic, {\em Across Traits} may inflate $\mivph_{vt}$ for the wrong
variant of the pair, thereby increasing FDR slightly for given $\xi$, but very
rarely increases $\mivph_{vt}$ by much for the causal variant; so power is not
increased.

%\vspace{3em}
\noindent
\underline{\bf Results for the case study}
%\vspace{1em}

\begin{figure}
\begin{center}
\includegraphics[scale=0.9]{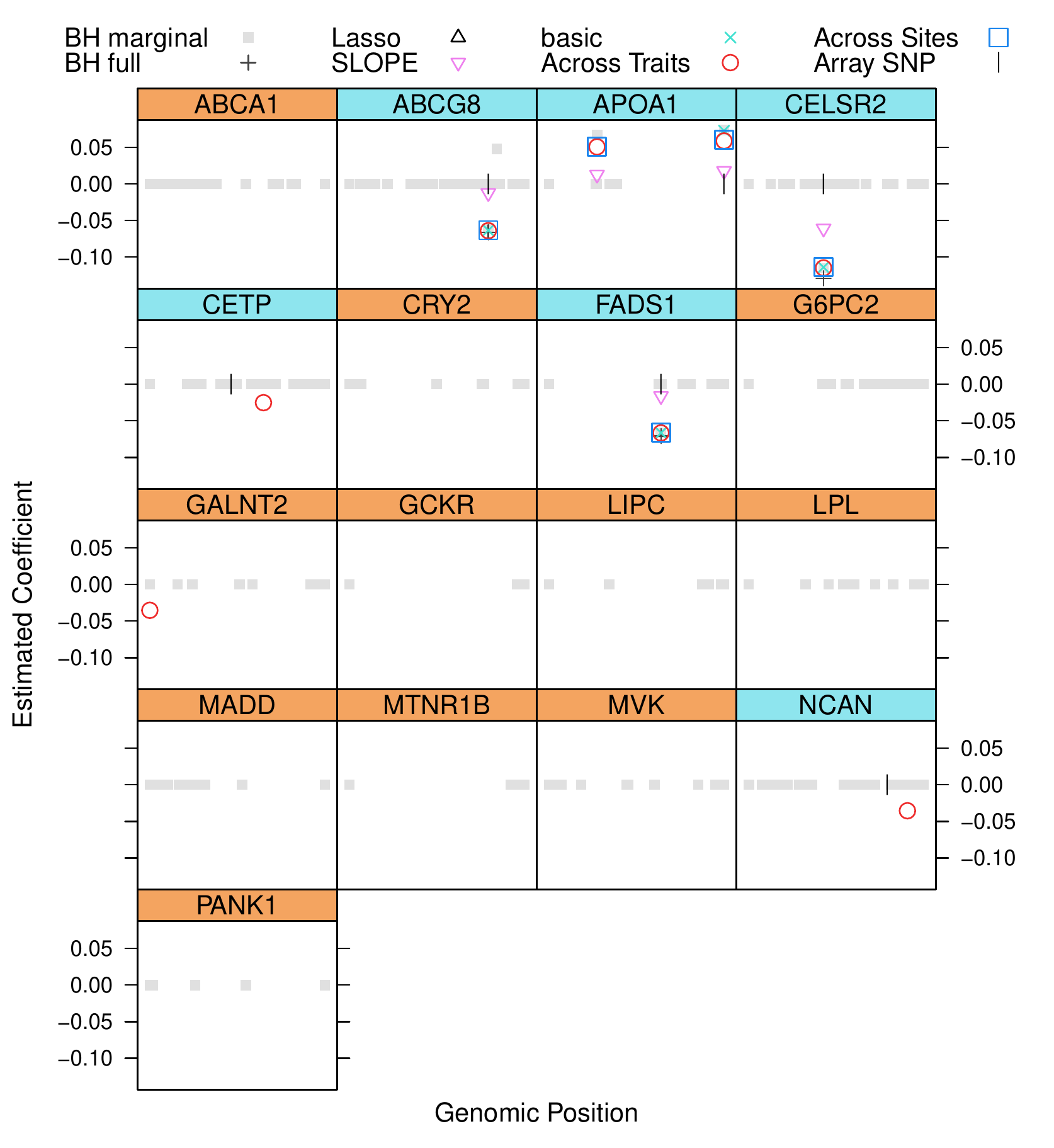}
\end{center}
\caption{\label{fig:LDL} Estimated variant effects on LDL. Each panel corresponds
to a locus,  the $x$-axis indicate the variant's genomic position and the
$y$-axis its regression coefficient (with the exception of BH marginal, only
nonzero coefficients are represented). The color code of the panel titles
indicates the presence/absence of prior evidence of association between the
locus and LDL (turquoise/orange, respectively). Model selection methods are
distinguished using plotting symbols, as indicated in the legend at the top. }
\end{figure}

\begin{figure}
\begin{center}
\includegraphics[scale=0.9]{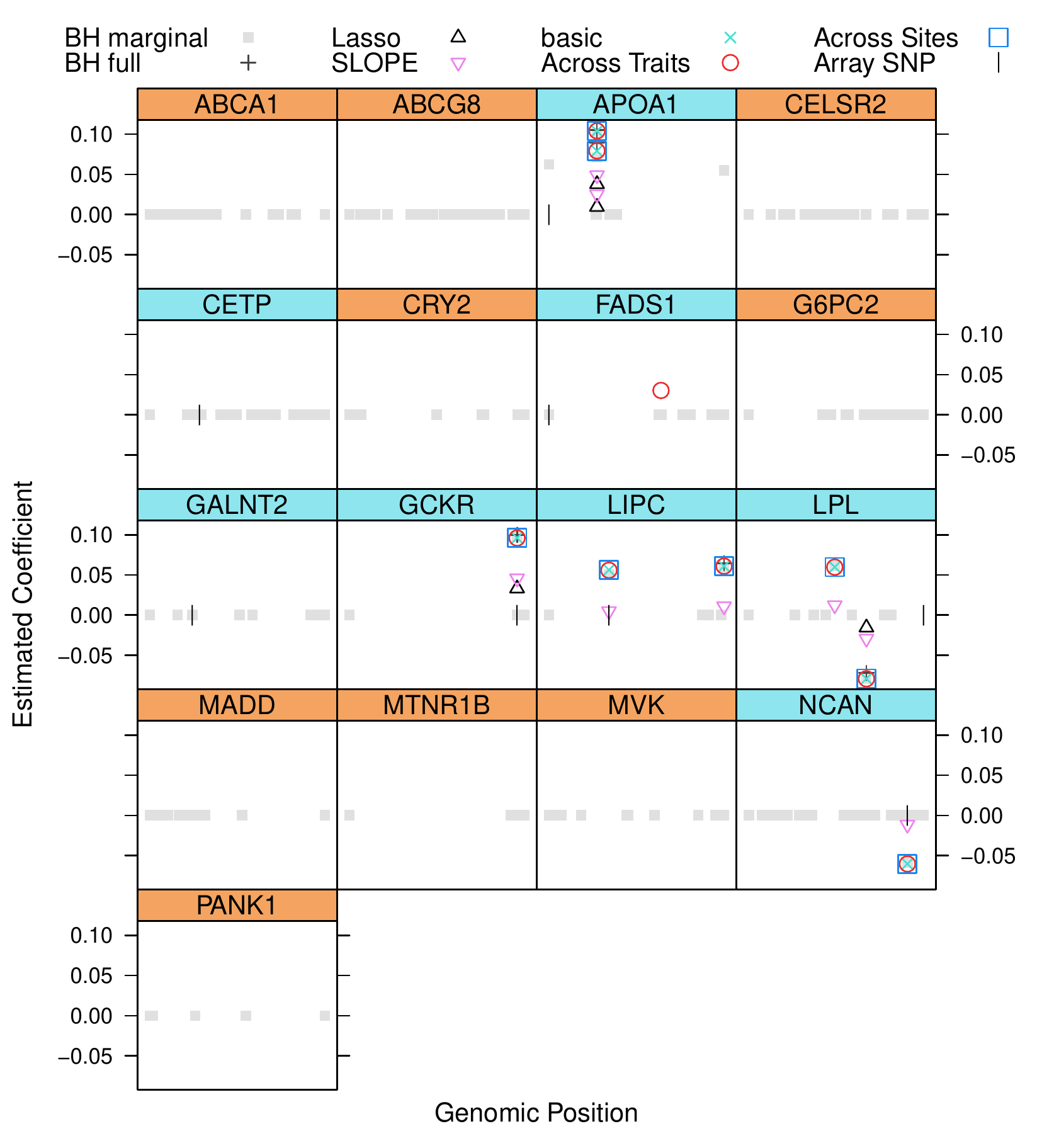}
\end{center}
\caption{\label{fig:TG} Estimated variant effects on TG. Each panel corresponds
to a locus,  the $x$-axis indicate the variant's genomic position and the
$y$-axis its regression coefficient (with the exception of BH marginal, only
nonzero coefficients are represented). The color code of the panel titles
indicates the presence/absence of prior evidence of association between the
locus and TG (turquoise/orange, respectively). Model selection methods are
distinguished using plotting symbols, as indicated in the legend at the top. }
\end{figure}

Figures~\ref{fig:LDL} and \ref{fig:TG} illustrate the model selection results
for LDL and TG, respectively, which are interpreted in the same way as the
figure for HDL in the main paper.  Tables~\ref{tbl:HDL}--\ref{tbl:TG} list the
methods that select each variant for association with HDL, LDL, and TG,
respectively, for variants selected by at least one method.  An {\tt x} in a
column indicates that the method selects that variant for that trait.
The column labeled `position' gives the position (from GRCh37) of the variant on
its chromosome.

\begin{table}
\caption{\label{tbl:HDL} Comparison of selection results for HDL.}
\center
\footnotesize
\begin{Verbatim}[frame=single]
                                        BH   BH                   Across Across
       variant  position  locus    MAF full marg Lasso SLOPE basic Trait Sites   PLOS
v_c9_107548661 107548661  ABCA1 0.0002   x   x     x     x     x             x    x
v_c9_107555091 107555091  ABCA1 0.0005       x           x     x             x    x
v_c9_107555452 107555452  ABCA1 0.0002   x   x     x     x     x      x      x    x
    rs76881554 107578620  ABCA1 0.0043                                       x     
     rs2066715 107588033  ABCA1 0.0551       x     x     x     x      x      x    x
     rs2066718 107589255  ABCA1 0.0148   x   x           x     x      x      x    x
   rs145183203 107646756  ABCA1 0.0029                                       x     
     rs2575875 107662494  ABCA1 0.2999       x     x     x     x      x      x    x
v_c9_107665945 107665945  ABCA1 0.0002                                       x     
    rs12805061 116553025  APOA1 0.2718       x           x                        x
      rs651821 116662579  APOA1 0.0845       x     x     x     x      x      x     
      rs646776 109818530 CELSR2 0.2171                                x            
    rs34216426  56913088   CETP 0.0002       x           x     x             x     
        rs5801  56913513   CETP 0.1818       x           x     x      x      x     
        rs5802  56919235   CETP 0.1194       x                                     
     rs3764261  56993324   CETP 0.2719   x   x     x     x     x      x      x    x
        rs5883  57007353   CETP 0.0401   x         x     x     x      x      x    x
        rs5880  57015091   CETP 0.0242   x   x     x     x     x      x      x    x
     rs2303790  57017292   CETP 0.0006   x               x     x      x      x    x
v_c16_57095439  57095439   CETP 0.0002       x           x                         
      rs509360  61548559  FADS1 0.3364                                            x
      rs174546  61569830  FADS1 0.4212                                x            
      rs611229 230324067 GALNT2 0.4077       x     x     x     x      x      x    x
     rs1532085  58683366   LIPC 0.4431   x   x     x     x     x      x      x    x
      rs261336  58742418   LIPC 0.2302   x   x     x     x     x      x      x     
    rs28933094  58855748   LIPC 0.0150   x   x     x     x     x      x      x    x
         rs268  19813529    LPL 0.0180   x   x     x     x     x      x      x    x
         rs328  19819724    LPL 0.0882   x   x     x     x     x      x      x    x
    rs10096633  19830921    LPL 0.0943                                            x
v_c11_47400044  47400044   MADD 0.0005   x   x     x     x     x             x     
     rs7946766  48004369   MADD 0.1965       x     x     x     x      x      x    x
    rs12314392 110010866    MVK 0.4033                                       x    x
\end{Verbatim}
\end{table}

\begin{table}
\caption{\label{tbl:LDL} Comparison of selection results for LDL.}
\center
\footnotesize
\begin{Verbatim}[frame=single]
                                        BH   BH                   Across Across
       variant  position  locus    MAF full marg Lasso SLOPE basic Trait Sites   PLOS
     rs6756629  44065090  ABCG8 0.0877   x   x           x     x      x      x    x
   rs145756111  44102302  ABCG8 0.0115       x                                    x
      rs651821 116662579  APOA1 0.0845       x           x            x      x    x
    rs11216267 116952392  APOA1 0.4578       x           x     x      x      x    x
      rs646776 109818530 CELSR2 0.2171   x   x           x     x      x      x    x
    rs12445698  56928216   CETP 0.1887                                            x
     rs3764261  56993324   CETP 0.2719                                x            
      rs174546  61569830  FADS1 0.4212   x   x           x     x      x      x    x
      rs611229 230324067 GALNT2 0.4077                                x            
    rs12610185  19721722   NCAN 0.0617                                            x
     rs2304130  19789528   NCAN 0.0610                                x            
\end{Verbatim}
\end{table}

For LDL and TG, no method identifies any rare variant, and the {\em Across
Sites} selections always agree with the majority of approaches.  The {\em Across
Traits} approach selects a few variants that are noteworthy.  For LDL, it
selects a variant in each of three loci---CETP, GALNT2, and NCAN---where no
other method identifies any signal.  In each case, the variant that the {\em
Across Traits} approach selects has a strong association with either HDL or
TG.  Two of these three loci have previous evidence of association with LDL.
For TG, the {\em Across Traits} approach selects a variant in FADS1, where no
other method selects any.  This is the same variant that caught our attention
when only the {\em Across Traits} approach selected it for HDL.  Again, this
locus has previous evidence of association with TG.

\begin{table}
\caption{\label{tbl:TG} Comparison of selection results for TG.}
\center
\footnotesize
\begin{Verbatim}[frame=single]
                                        BH   BH                   Across Across
       variant  position  locus    MAF full marg Lasso SLOPE basic Trait Sites   PLOS
    rs12805061 116553025  APOA1 0.2718       x                                    x
     rs2266788 116660686  APOA1 0.0891                                            x
     rs3135506 116662407  APOA1 0.0597   x   x     x     x     x      x      x    x
      rs651821 116662579  APOA1 0.0845   x   x     x     x     x      x      x     
    rs11216267 116952392  APOA1 0.4578       x                                     
     rs1561140  56864398   CETP 0.4762                                            x
     rs6591657  61434532  FADS1 0.1506                                            x
      rs174546  61569830  FADS1 0.4212                                x            
     rs4846930 230346829 GALNT2 0.4050                                            x
     rs1260326  27730940   GCKR 0.3566   x   x     x     x     x      x      x    x
      rs261336  58742418   LIPC 0.2302       x           x     x      x      x    x
    rs28933094  58855748   LIPC 0.0150   x   x           x     x      x      x    x
         rs268  19813529    LPL 0.0180       x           x     x      x      x    x
         rs328  19819724    LPL 0.0882   x   x     x     x     x      x      x    x
    rs10096633  19830921    LPL 0.0943                                            x
     rs2304130  19789528   NCAN 0.0610       x           x     x      x      x    x
\end{Verbatim}
\end{table}

To conclude, we note that applying BH to each phenotype separately gives
different results in some cases.  When applied to the p-values from the full
model for HDL only, BH selects four variants that it does not select when
applied to all traits simultaneously: {\it rs2066715} in ABCA1,
{\it rs651821} in APOA1, {\it rs34216426} in  CETP, and {\it rs611229} in
GALNT2.  All of these are also selected for HDL by most other methods.
When applied to marginal p-values for HDL only, BH selects three variants that
it does not select when applied to all traits simultaneously: {\it rs62136410}
in ABCG8, {\it rs2303790} in CETP, and {\it rs11988} in MADD.  Most methods
select {\it rs2303790} for HDL, but no other methods select the other two of
these; and ABCG8 does not even have prior evidence of association to HDL.  The
final difference arising from applying BH to traits separately is that it does
not select {\it rs145756111} in ABCG8 for association with LDL when using
marginal p-values, which brings it into agreement with the other methods on this
variant.

%\vspace{3em}
\noindent
\underline{\bf Effect of pruning variants}
%\vspace{1em}

% Alas, LaTeX command cannot contain a number.
\newcommand{\bXa}{\bX^{(3)}}
\newcommand{\bXb}{\bX^{(5)}}
\newcommand{\bXc}{\bX^{(7)}}
\newcommand{\bXd}{\bX^{(9)}}
For the case study, we pruned the set of variants so that the maximum of the
absolute values of the pairwise correlations is 0.3 in order to facilitate
comparison of the selection results by different methods.  We will now describe
some experiments we performed with the actual data to investigate the effects of
this pruning.  First, we compare our selection results presented in the main
text and above to the results when using $\bX$ obtained by setting $\cmax$ in
Figure~\ref{fig:filter} equal to 0.5, 0.7 and 0.9.  For clarity, let $\bXa$,
$\bXb$, $\bXc$ and $\bXd$ denote the different versions of $\bX$.
After 5,000,000 MCMC
iterations for each of four chains, over 99\% of $\Delta\mivph_{vt}$ and of
$\Delta\overline{\ivar}_v$ or $\Delta\overline{\igrp}_g$ are less than 0.05 for
all levels of pruning, so convergence still seems acceptable.

The values of $\mivph_{vt}$ usually do not change very much for different
versions of $\bX$; and when they do, they almost always decrease.  More
precisely, much less than 1\% of the values of $\mivph_{vt}$ {\em increase} by
more than 0.05.  For given prior and trait, the percentage of values that {\em
change} by more than 0.05 is less than 3\% between $\bXa$ and $\bXb$ and 1\%
between $\bXb$ and $\bXc$ and between $\bXc$ and $\bXd$.

Table~\ref{tbl:prunesel} compares the total number of variants selected
by our Bayesian priors for each of the pruning levels; they still did not select
any variants that are in a locus (CRY2, G6PC2, MTNR1B, and PANK1) lacking any
prior evidence of association to lipid traits.  The table also shows the number
of variants selected by applying BH to the p-values from $\bXa$ and $\bXd$.
\begin{table}
\caption{\label{tbl:prunesel}  Comparison of selection results from BH and
Bayesian methods applied to different levels of variant pruning.
The columns labeled $R$ give the number of variants selected, and
the columns labeled $\hbfdr$ report the Bayesian FDR.
The selection threshold for the Bayesian methods was $\xi=0.7$.
For BH, $\alpha=0.05$.}
\centering
\begin{tabular}{l||rr|rr|rr}
& \multicolumn{2}{c|}{HDL} & \multicolumn{2}{c|}{LDL} & \multicolumn{2}{c}{TG}\\
\hline
\hline
prior &  $R$&$\hbfdr$ & $R$&$\hbfdr$ &  $R$&$\hbfdr$\\
\hline  
\hline  
\multicolumn{7}{c}{maximum correlation 0.3}\\
\hline  
basic                  & 21& 0.046 &  4& 0.012 &  8& 0.011\\
{\em Across Traits}    & 19& 0.084 &  8& 0.059 &  9& 0.029\\
{\em Across Sites}     & 25& 0.064 &  5& 0.063 &  8& 0.007\\
BH full                & 13&       &  3&       &  5&      \\
BH marginal            & 22&       &  6&       & 10&      \\
\hline  
\multicolumn{7}{c}{maximum correlation 0.5}\\
\hline  
basic                  & 21& 0.068 &  4& 0.014 &  9& 0.024\\
{\em Across Traits}    & 18& 0.091 &  8& 0.060 & 10& 0.044\\
{\em Across Sites}     & 25& 0.076 &  4& 0.012 &  9& 0.020\\
\hline  
\multicolumn{7}{c}{maximum correlation 0.7}\\
\hline  
basic                  & 20& 0.067 &  4& 0.014 &  9& 0.033\\
{\em Across Traits}    & 17& 0.081 &  8& 0.057 &  9& 0.022\\
{\em Across Sites}     & 24& 0.071 &  4& 0.013 &  9& 0.029\\
\hline  
\multicolumn{7}{c}{maximum correlation 0.9}\\
\hline  
basic                  & 19& 0.073 &  4& 0.038 &  9& 0.050\\
{\em Across Traits}    & 15& 0.071 &  8& 0.090 &  9& 0.048\\
{\em Across Sites}     & 22& 0.071 &  4& 0.037 &  9& 0.045\\
BH full                &  6&       &  2&       &  4&      \\
BH marginal            & 32&       & 13&       & 15&
\end{tabular}
\end{table}
For a more detailed look at the differences, Table~\ref{tbl:prunedetail} lists
all cases in which a variant would be selected for one version of $\bX$ but not
another.
\begin{table}
\caption{\label{tbl:prunedetail} Cases in which a variant would be selected for
one version of $\bX$ but not another.  Entries show $\mivph_{vt}$; ``NA'' means
that the variant is not in that version of $\bX$, while blanks mean that the
variant would not be selected for any $\bX$.}
\centering
\begin{tabular}{l|rrrr|rrrr|rrrr}
        & \multicolumn{4}{c|}{none}
        & \multicolumn{4}{c|}{\emph{Across Traits}}
        & \multicolumn{4}{c}{\emph{Across Sites}}\\
\hline
\hline
max correlation & 0.3  & 0.5  & 0.7 & 0.9 & 0.3  & 0.5  & 0.7 & 0.9 & 0.3  & 0.5
& 0.7 & 0.9\\
\hline
\multicolumn{13}{c}{HDL}\\
\hline
\it rs2066715& 0.92& 0.89& 0.23& 0.27& 0.78& 0.75& 0.17& 0.22
                            & 0.92& 0.89& 0.13& 0.13\\
\it rs2853579&   NA&   NA& 0.75& 0.70&   NA&   NA& 0.75& 0.69
                            &   NA&   NA& 0.88& 0.87\\
\it rs140547417&   &     &     & &     &     &     & &   NA& 0.73& 0.74& 0.66\\
\it rs2575875& 0.92& 0.73& 0.40& 0.40& 0.79& 0.66& 0.39& 0.41
                            & 0.94& 0.82& 0.46& 0.47\\
\it rs12314392 &   &     &     & &     &     &     & & 0.73& 0.54& 0.53& 0.19\\
\it rs7946766& 0.99& 0.97& 0.97& 0.56& 0.98& 0.97& 0.97& 0.52
                            & 1.00& 0.98& 0.98& 0.65\\
\it rs174546 &     &     &     & & 0.72& 0.70& 0.62& 0.58& \\
\hline
\multicolumn{13}{c}{LDL}\\
\hline
\it rs651821 &  &  &  &    & \multicolumn{4}{c|}{always \textgreater0.99}
                            &0.73& 0.66& 0.63& 0.64\\
\hline
\multicolumn{13}{c}{TG}\\
\hline
\it v\_c9\_107544165& NA& 0.86& 0.85& 0.83& NA& 0.83& 0.82& 0.79& NA& 0.87& 0.87& 0.87\\
\it rs174546   &   &     &     &    & 0.74& 0.74& 0.68& 0.64
\end{tabular}
\end{table}
These differences are explained in detail at the end of this section, but for
now we categorize them as follows:
(1)~one variant ({\it rs2066715}) is replaced in the selections by another
strongly correlated variant ({\it rs2853579}) when it enters $\bXc$;
(2)~two rare variants are not in $\bXa$ but are selected (or close to the
threshold) for the other versions of $\bX$, at least in part because one of the
2--3 subjects with the minor allele has an extreme value for the trait;
(3)~two variants whose $\mivph_{vt}$ is barely above the selection threshold for
some trait and prior with $\bXa$ but then decreases by as much as 0.15 as
correlated variants enter $\bX$ with $\mivph_{vt} < 0.1$; and
(4)~three variants that are selected for $\bXa$ but their $\mivph_{vt}$
decreases by as much as 0.5 as correlated variants enter $\bX$ with
$\mivph_{vt} > 0.15$.

To summarize, the different versions of $\bX$ considered here do not have much
effect on the selections made when computing $\mivph_{vt}$ from MCMC across all
loci simultaneously.  Very few of the variants that were left out of $\bXa$ are
selected when they are included in a different version of $\bX$.  In fact,
adding correlated variants to $\bX$ is more likely to result in a slight
reduction in the number of selected variants.
In contrast, applying BH to the p-values has a much
greater effect on the selections between $\bXa$ and $\bXb$, with marginal
p-values resulting in many more selections and p-values from the full model
resulting in many fewer selections.

Variants with sufficiently strong correlations could, however, lead to
convergence problems in our MCMC.  To allow arbitrary correlations,
we also compute the posterior distribution exactly instead of using MCMC.
To achieve this, we make the following simplifications as in
\cite{SerS07,HetE14,CheL15}:
\begin{itemize}
\item Specify $\tau$.  We choose $\tau=0.05$ because the mean of the sampled
values of $\tau$ in the MCMC was always between 0.04 and 0.06 for all versions
of $\bX$ with the actual phenotype data.
\item Assume the posterior density is zero if $\abs{\bvph}\geq K$, where $K$ is
typically no more than six but may be as small as one.
\end{itemize}
Even requiring $\abs{\bvph} \leq 4$, however, would still leave far too many
subsets if we were to consider all 1302 variants simultaneously, so we only
consider one locus at a time.  In fact, we consider only two loci: ABCA1 and
CETP, both with trait HDL, because these show stronger evidence of influencing
the trait via multiple variants.  The locus ABCA1 has 58 variants, so we
compute the posterior for $\abs{\bvph}\leq6$; whereas the locus CETP has 98
variants, so we require $\abs{\bvph}\leq5$.  Finally, we only tried this with
our basic model.

In addition to estimating the marginal posterior expected values as
$\expect[\ivph_v|\by]\approx\sum_{\bvph:\ivph_v=1}
f_{\bvph,\tau}(\bvph,\tau=0.05|\by)$, we also consider two other approaches to
selection.  One is simply to choose $\bvph$ with the largest value of
$f_{\bvph,\tau}(\bvph,\tau=0.05|\by)$.  The other is to use the confidence set
proposed by \cite{HetE14}, described as follows.  The probability that a set $\mathcal{S}$ of variants
contains all causal variants is $\mathcal{P}(\mathcal{S})\equiv
\sum_{\bvph:\ivph_v=0\text{ for }v\in\mathcal{S}}
         f_{\bvph,\tau}(\bvph,\tau=0.05|\by)$.  Ideally, one would choose the
smallest $\mathcal{S}$ such that $\mathcal{P}(\mathcal{S})$ is above a specified
threshold, but this is not computationally feasible.  Instead, we use the
stepwise selection process in \cite{HetE14}.  At each step, the variant that
increases $\mathcal{P}(\mathcal{S})$ the most is added to $\mathcal{S}$.
\citep{HetE14} use the stopping criterion $\mathcal{P}(\mathcal{S})\geq0.95$,
but this results in a very large number of selections for our data when the
maximum value of $\abs{\bvph}$ is four or greater.  Instead, we use threshold
0.7.

The 70\%-confidence set contains all the variants selected by any of the
Bayesian methods, so Table~\ref{tbl:exactcomp} lists the variants in the order
in which they enter the 70\%-confidence set.  The selections from BH applied to
the p-values for only the one locus and the HDL are also shown, although BH
applied to the marginal p-values for locus CETP selects nine additional variants
that are not shown.  \cite{CheL15} give a good argument for
preferring selections based on marginal posterior expectations over the
confidence set if there is more than one causal variant, so we do not discuss
the confidence sets further here.  BH applied to marginal p-values is probably
also selecting too many variants.  The other methods highlight many of the
same variants.  Overall, for 90\% of the variants in $\bXd$, the value of
$\mivph_{vt}$ computed by MCMC is within 0.01 of the exact posterior marginals
without pruning, and the latter value is almost always less than the former,
which again suggests that pruning did not result in many missed discoveries.
One of the largest differences between these two values is for {\it rs2853579},
which---as noted above in connection with Table~\ref{tbl:prunedetail}---is
strongly correlated with {\it rs2066715}.
Furthermore, for locus ABCA1, the set $\bvph$ of variants with the largest
posterior density is equal to the variants selected by using $\mivph_{vt}$ with
$\bXa$.  For the locus CETP, these two sets cannot be the same because
$\mivph_{vt}$ selects six variants, while we had to assume $\abs{\bvph}\leq5$ in
order to compute the exact distribution; additionally, the mode of the posterior
density includes one very rare variant (MAF 0.0002) that is in $\bXa$ but that
the basic prior does not select.

\begin{sidewaystable}
\footnotesize
\caption{\label{tbl:exactcomp} Comparison of selection methods to see effect of
using all variants.  These tables list all variants in the 70\%-confidence set
in the order in which they would be added to this set.
The variant names in bold face have $\mivph_{vt} > 0.7$ from the MCMC with
$\bXa$, while those preceded by an asterisk are \emph{not} in $\bXa$.
In the columns for $\mivph_{vt}$ from MCMC, a blank means that $\mivph_{vt} <
0.2$, and NA means that the variant is not in that $\bXd$.
In the columns for the exact marginal, a blank means that the value is less than
0.2.}
\vspace{0.2cm}
\centering
\begin{tabular}{l||c|cccc}
\multicolumn{6}{c}{locus ABCA1 with response HDL}\\
\hline
\hline
       & $\bXd$      & \multicolumn{4}{c}{all variants in locus}\\
\cline{3-6}
       & MCMC        & exact   &argmax$_{\bvph}$    & BH & BH\\
variant&$\mivph_{vt}$&marginal& $f(\bvph|\tau,\by)$&full&marginal\\
\hline
\bf  rs2575875       & 0.40 & 0.48& yes &     & yes\\
\bf v\_c9\_107555452 & 0.99 & 0.97& yes & yes & yes\\
\bf v\_c9\_107548661 & 0.85 & 0.81& yes & yes & yes\\
\bf  rs2066718       & 0.93 & 0.73& yes & yes & yes\\
\bf  rs2066715       & 0.27 & 0.28& yes &     & yes\\
\bf v\_c9\_107555091 & 0.87 & 0.62& yes & yes & yes\\
*rs1800978           & 0.33 & 0.26&     &     & yes\\
*rs2853579           & 0.70 & 0.32&     &     & yes\\
*rs2740486           & 0.36 & 0.25&     &     & yes\\
    rs76881554       & 0.46 &     &     &     &    \\
*rs2066714           &  NA  &     &     &     & yes\\
*rs2230806           &      &     &     &     &    \\
*rs2066716           &      &     &     &     & yes\\
\hline
\multicolumn{6}{c}{end of 50\%-confidence set}\\
\hline
*rs2230805           & NA   &     &         & \\
     rs2230808       & 0.24 &     &         & \\
*v\_c9\_107544165    &      &     &         & \\
    rs41437944       &      &     &         & \\
   rs145183203       &      &     &         & \\
     rs9282537       &      &     &          
\end{tabular}
\hspace{2em}
\begin{tabular}{l||c|cccc}
\multicolumn{6}{c}{locus CETP with response HDL}\\
\hline
\hline
       & $\bXd$      & \multicolumn{4}{c}{all variants in locus}\\
\cline{3-6}
       & MCMC        & exact   &argmax$_{\bvph}$    & BH & BH\\
variant&$\mivph_{vt}$&marginal& $f(\bvph|\tau,\by)$&full&marginal\\
\hline
\bf  rs3764261   & 1.00 & 1.00& yes  &yes & yes\\
\bf     rs5883   & 1.00 & 0.98& yes  &yes & yes\\
\bf     rs5880   & 0.95 & 0.92& yes  &     & yes\\
\bf rs34216426   & 0.86 & 0.62& yes  &yes & yes\\
v\_c16\_57095439 & 0.46 & 0.28& yes  &yes & yes\\
\bf  rs2303790   & 0.92 & 0.29&      &yes & yes\\
\bf     rs5801   & 0.86 & 0.26&      &     & yes\\
\hline                                       
\multicolumn{6}{c}{end of 50\%-confidence set}\\
\hline
*v\_c16\_56926936&      &     &      &     &     \\
   rs140440847   &      &     &      &     & yes\\
*rs140547417     &      &     &      &     &     \\
v\_c16\_57117215 & 0.22 &     &      &     &     \\
*rs12445698      &      &     &      &     & yes\\
     rs3751705   &      &     &      &yes  &     
\end{tabular}                                 
\end{sidewaystable}

For completeness, the detailed descriptions of the differences in
Table~\ref{tbl:prunedetail} are as follows:
\begin{description}
\item[{\it rs2066715} and {\it rs2853579},] which are both in ABCA1, are in the
table for all three priors applied to response HDL.  Their correlation is just
below 0.7, and $\mivph_{vt}$ for the former decreases drastically when the
latter is in $\bX$ and is selected instead---although $\mivph_{vt}$ for {\it
rs2853579} falls just below the selection threshold with {\em Across Traits}
applied to $\bXd$.
The former is a missense variant predicted to be benign; whereas the
latter is a synonymous variant.
\item[\it rs140547417] is not in $\bXa$, has $\mivph_{vt}$ barely above the
selection threshold for $\bXb$ and $\bXc$ when {\em Across Sites} is applied to
HDL, then drops slightly below the threshold for $\bXd$.
Its minor allele occurs in only three subjects, one of which
has the 32nd largest value of HDL and another is in the 15th-percentile.
Furthermore, this variant is in gene CETP, and its group includes a variant with
$\mivph_{vt}\approx0.8$ and one with $\mivph_{vt}\approx0.3$.
\item[\it rs2575875] is strongly correlated with a new variant in $\bXb$
(with $\mivph_{vt}=0.24$) and with another new variant in $\bXc$---at which
point all three variants have $\mivph_{vt}$ in 0.3--0.4.
%It has correlation 0.45 with {\it rs2740486} and correlation 0.52 with {\it
%rs1800978}.  When {\it rs2575875} and {\it rs2740486} are in both in $\bX$,
%then $\mivph_{vt}$ for {\it rs2575875} decreases slightly while $\mivph_{vt}$
%is about 0.2 for {\it rs2740486}.  When all three of these variants are in
%$\bX$, then each has $\mivph_{vt}$ roughly 0.3--0.4.
%The variant {\it rs2575875} is an Array SNP for HDL and {\it rs2740486} is an
%Array SNP for total cholesterol; whereas {\it rs1800978} is in the
%UTR~$5^\prime$.
\item[\it rs12314392] has $\mivph_{vt}$ barely above the selection threshold
when {\em Across Sites} is applied to $\bXa$.  It has correlation 0.3--0.4 with
three variants that are in $\bXb$.  Two of these variants have
$\mivph_{vt} < 0.05$, while the other has $\mivph_{vt}=0.2$; this is easily
enough for {\it rs12314392} to drop below the selection threshold.  None of the
new variants in $\bXc$ are strongly correlated with {\it rs12314392}, but four
of the new variants in $\bXd$ have correlations 0.43--0.84 with {\it
rs12314392}.  Three of these have $\mivph_{vt} < 0.1$; but the one with
correlation 0.75 has $\mivph_{vt} = 0.60$, causing $\mivph_{vt}$ for {\it
rs12314392} to decrease even more.
\item[\it rs7946766] has $\mivph_{vt} > 0.96$ for HDL with all three priors
until several variants strongly correlated with it are added to $\bXd$,
including two that have $\mivph_{vt} \approx 0.3$.
\item[\it v\_c9\_107544165] has minor allele that occurs in only two subjects, one of
which has the 2nd largest value of HDL and the other is in the 14th-percentile.
It is in the UTR~$3^\prime$ of ABCA1, which means it is in a group by itself in
the {\em Across Sites} prior.
\item[{\it rs174546} and {\it rs651821}] both have $\mivph_{vt}$ between 0.7 and
0.75 for $\bXa$, but it decreases as more variables are added to $\bX$ until it
is between 0.58 and 0.65 for $\bXd$; in other words, these variants are near the
threshold for selection for all versions of $\bX$.
\end{description}

\end{document}